\documentclass[11pt,a4paper]{article} 
\usepackage{jheppub}

\usepackage{multirow, graphicx,amssymb,url,mathrsfs,amsmath}
\usepackage{wrapfig,boxedminipage,setspace,subfigure,epsfig}
\usepackage{amsxtra,amstext,latexsym,dsfont,amsfonts}
\def\a  {\alpha}                
               
\def\e  {\epsilon}

   \def\w  {\omega}

 \newcommand{\call}{\mbox{${\cal L}$}}
 \newcommand{\caln}{\mbox{${\cal N}$}}

\def\IR{{\hbox{{\rm I}\kern-.2em\hbox{\rm R}}}}
\def\IB{{\hbox{{\rm I}\kern-.2em\hbox{\rm B}}}}
\def\IN{{\hbox{{\rm I}\kern-.2em\hbox{\rm N}}}}
\def\IC{\,\,{\hbox{{\rm I}\kern-.59em\hbox{\bf C}}}}
\def\IZ{{\hbox{{\rm Z}\kern-.4em\hbox{\rm Z}}}}
\def\IP{{\hbox{{\rm I}\kern-.2em\hbox{\rm P}}}}
\def\IH{{\hbox{{\rm I}\kern-.4em\hbox{\rm H}}}}
\def\ID{{\hbox{{\rm I}\kern-.2em\hbox{\rm D}}}}

\def\be{\begin{equation}}
\def\ee{\end{equation}}
\def\ba{\begin{eqnarray}}
\def\ea{\end{eqnarray}}

\def\ra{\rightarrow}  
\def\lra{\Longrightarrow}

\def\nn{\nonumber}
\def\ea{{\it et al}. }

\newcommand{\wt}{\widetilde}

\newcommand{\beq}{\begin{equation}}
\newcommand{\eeq}{\end{equation}}
\newcommand{\bea}{\begin{eqnarray}}
\newcommand{\eea}{\end{eqnarray}}

\title{E, B, $\mu$, T Phase Structure  of the D3/D7 Holographic Dual}

\author{Nick Evans,}
\author{Astrid Gebauer,}
\author{and Keun-Young Kim}

\emailAdd{evans@soton.ac.uk}
\emailAdd{ag806@soton.ac.uk}
\emailAdd{k.kim@soton.ac.uk}

\affiliation{ School of Physics and Astronomy, University of
Southampton, \\ Southampton, SO17 1BJ, UK.
}

\abstract{ The large $N$ ${\cal N}$=4 gauge theory with quenched
${\cal N}$=2 quark matter displays chiral symmetry breaking in the
presence of a magnetic field. We previously studied the
temperature and chemical potential phase structure of this theory 
in the grand canonical ensemble 
- here we, in addition, include the effect of an electric field
which acts to counter chiral symmetry breaking 
by disassociating mesons. 
We compute using the gravity dual based on the D3/probe-D7
brane system. The theory displays two transition at one of which
chiral symmetry is restored. 
At the other transition density switches on, the mesons of the theory become unstable
and a current forms, making it a conductor-insulator transition.
Through the temperature,
electric field, chemical potential volume (at fixed magnetic field
parallel to the electric field) these transitions can coincide or
separate at critical points, and be first order or second order.
We map out this full phase structure which provides varied
computable examples relevant to strongly coupled gauge theories
and potentially condensed matter systems.}

\keywords{Gauge/Gravity duality}

\subheader{SHEP-11-05}

\begin{document}

\maketitle

%\preprint{
%\begin{minipage}[t]{3in}
%\begin{flushright} SHEP-11-05
%\\[30pt]
%\hphantom{.}
%\end{flushright}
%\end{minipage}
%}

\section{Introduction and summary}

Holographic techniques~\cite{Malda,Witten:1998qj,Gubser:1998bc}
have begun to allow us to study the phase structure of strongly
coupled gauge theories. 
Such studies are potentially of interest
for QCD, more exotic gauge theories and even to condensed matter
systems. Although it is somewhat hard to dial a particular gauge
theory, we instead work in tractable models that lie close to
${\cal N}=4$ Super Yang-Mills theory and hope that these gauge
theories show similar behaviours to realistic cases.

A simple model to study is the large $N$ ${\cal N}=2$ gauge theory made from
${\cal N}=4$ SYM plus a small number of quark hypermultiplets 
\cite{Karch,Polchinski,Bertolini:2001qa,Mateos,Erdmenger:2007cm}. In
the limit $N_f \ll N$ quark loops are quenched which on the gravity
side implies that one can study probe D7 (or D5) branes in the AdS
background of $N$ D3 branes. In the presence of a magnetic field ($B$) the theory dynamically
generates a quark condensate that breaks a chiral $U(1)$ global
symmetry \cite{Johnson1}. This dynamics may be relevant to condensed matter
systems~\cite{Evans2, Jensen, Evans3} but one can also study it as a loose analogue of QCD-like
gauge theories \cite{Evans1}. 
Of course in those theories the running coupling
generates a scale and the chiral symmetry breaking, but the hard
magnetic field can be thought of as the analogue of the scale
$\Lambda_{QCD}$ which breaks the conformal symmetry and ``allows"
the strong dynamics to generate a condensate ($c$).

\begin{figure}[]
\centering
  \subfigure[$T$-$\mu$ phase diagram. The inset diagrams are representative  probe brane embeddings (dotted lines), 
    where a black disk represents a black hole.]
   {\includegraphics[width=7cm]{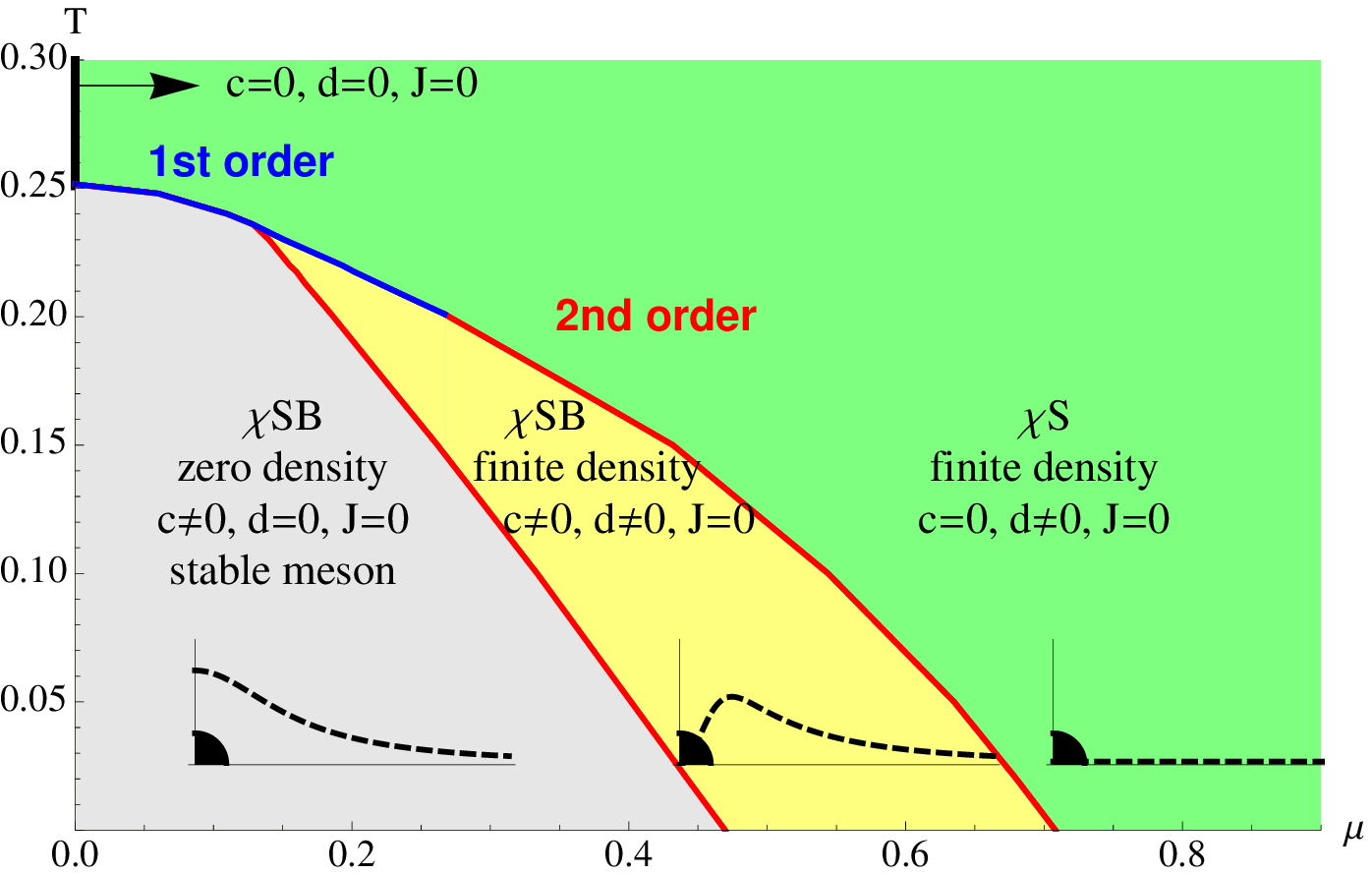} \label{Tvsmu}} \quad
  \subfigure[$T$-$E$ phase diagram. The inset diagrams are representative probe brane embeddings (dotted lines), 
    where a red arc represents a singular shell.]
   {\includegraphics[width=7cm]{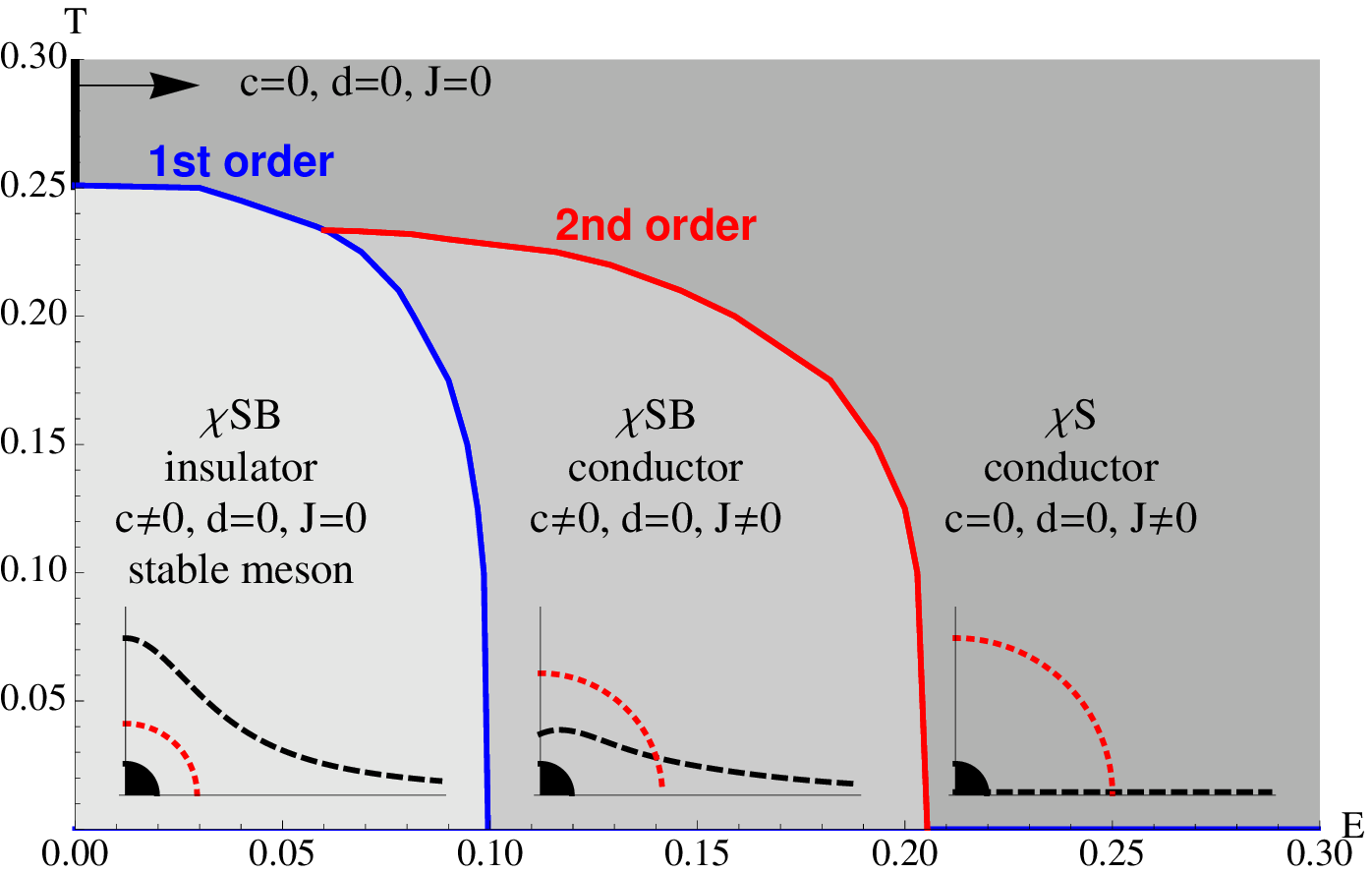} \label{TvsE} }
   \subfigure[$\mu$-$E$ phase diagram. The $\mu$($E$) axis agrees with the  $\mu$($E$) axis of Fig \ref{Tvsmu}(\ref{TvsE}) and are
   shown by the same color. Thus embeddings are the same as \ref{Tvsmu} and \ref{TvsE} without the black hole.
    ]
   {\includegraphics[width=7cm]{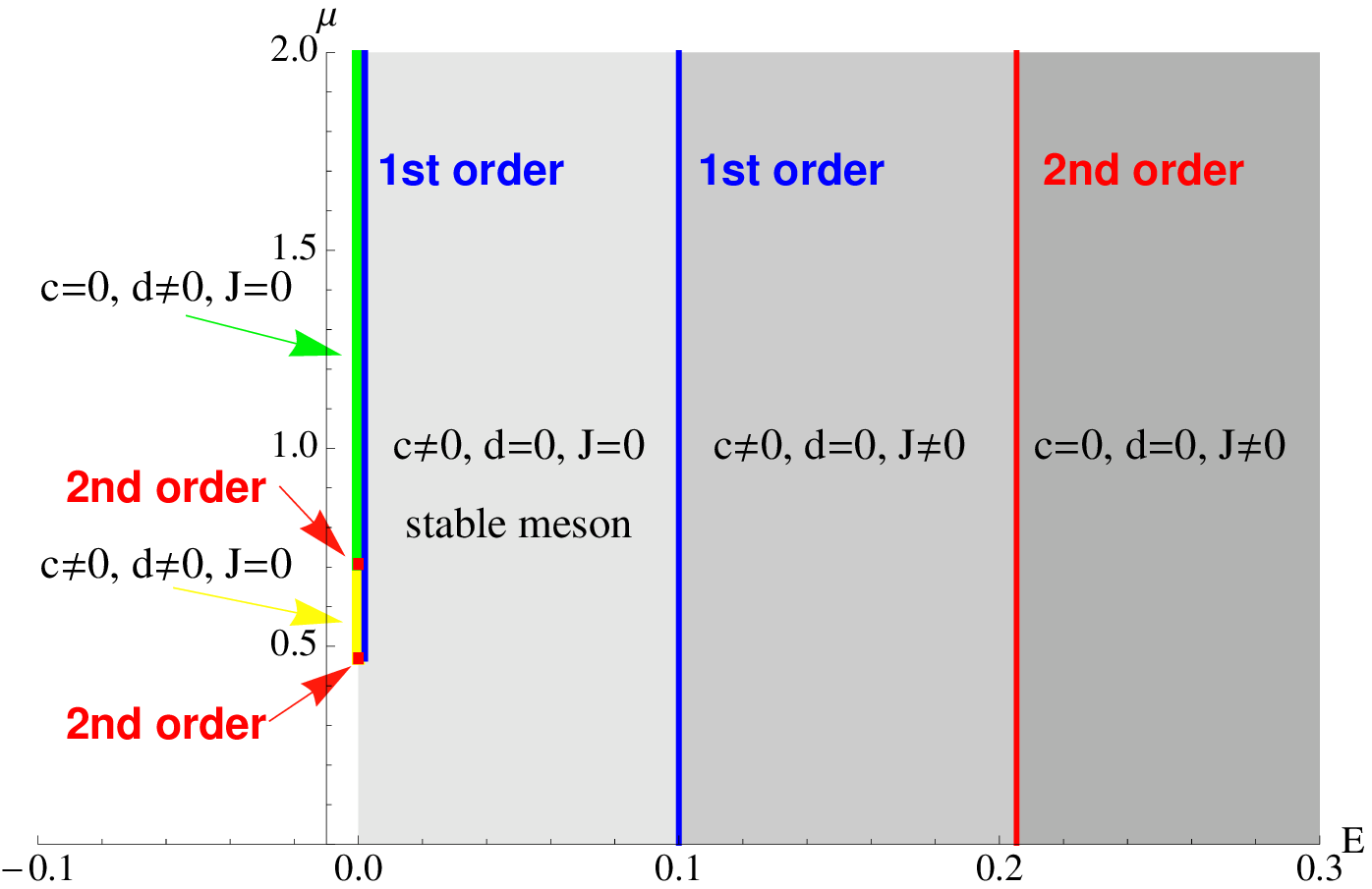} \label{muvsE} } \quad
  \subfigure[$E$-$\mu$-$T$ phase diagram. 
            The curves show the phase boundaries at fixed $T$ 
            ($T=0$, $0.15$, $0.21$, $0.232$, $0.24$ from the bottom).
            See Fig \ref{FixedT} for more detail.]
   {\includegraphics[width=7cm]{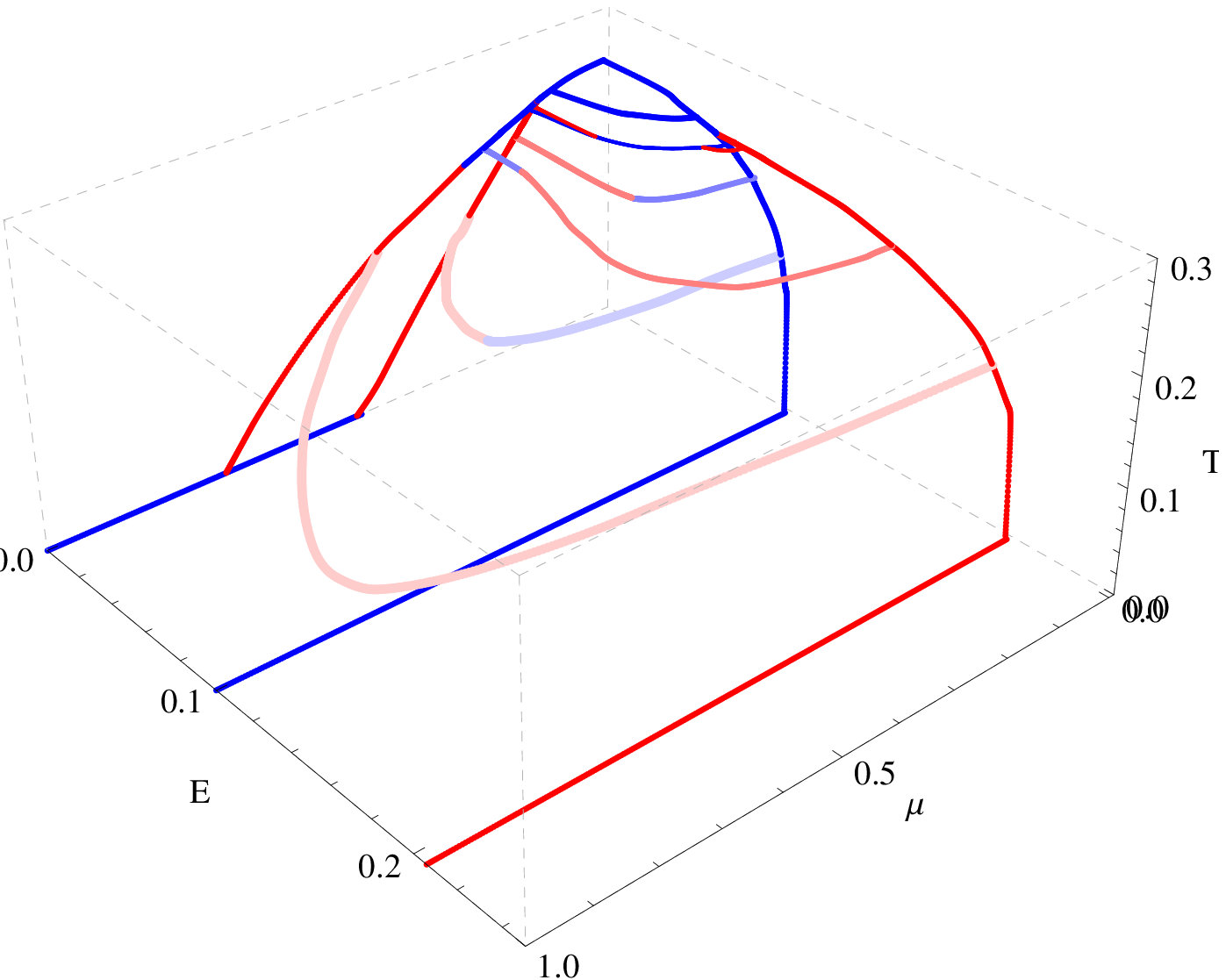}  \label{3D}}
  \caption{{The phase diagrams of the massless ${\cal N}=2$ gauge theory with
  a magnetic field. First order transitions are shown in blue,
  second order transitions in red. The temperature is controlled by the parameter $T$,
  chemical potential by $\mu$ and electric field by $E$.}
           }
\end{figure}

In \cite{Evans1} we computed the phase diagram for the theory with massless quarks, in a magnetic
field, in the
temperature chemical potential plane ($T$-$\mu$ plane) 
which we show in Fig \ref{Tvsmu}.
It has considerable structure. Whilst the magnetic field favours chiral
symmetry breaking the temperature \cite{Babington} and chemical potential 
\cite{Myers1} favour the chirally symmetric phase.
This tension leads to interesting phase structure.
There are three phases - the high
$T$, $\mu$ phase with the symmetry restored ($\chi$S), finite density ($d \ne 0$) and no stable mesonic
states; an intermediate phase in which chiral symmetry is broken ($\chi$SB), finite density ($d\ne0$), 
but mesons are still unstable; and finally a low $T$, $\mu$ phase
with broken chiral symmetry ($\chi$SB), no density ($d=0$) and stable mesons.  
On the gravity side these phases are described by 
a flat D7 embedding, an embedding that spikes on to the black hole horizon,
and a solution that lies outside the horizon (See the insets in Fig \ref{Tvsmu} for schematic plots.). 
These phases are linked by transition lines that are in places first order (blue lines) and in
places second order (red lines) - critical points link these behaviours.
Time dependent aspects of these transitions have also been studied in \cite{Evans4,Grosse:2007ty,Guralnik:2011xz,Evans5}. 

In this paper we wish to study how robust the phase structure is
to changes of parameters. In particular we will introduce the
additional parameter of an electric field ($E$) parallel to the magnetic
field. Note this case is simpler than when the $E$ and $B$ fields are
perpendicular because no Hall current forms.  We do allow for the
induced current in the direction of the electric field. The effect
on the theory of an electric field has been studied previously for probe branes in
\cite{Andy1, Hall, Erdmenger1, Veselin1, Andy2, Mas, Andy3}. Although the electric field continuously
acts on the quarks an equilibrium configuration is nevertheless
reached where energy is being dissipated to the bulk ${\cal N}=4$ plasma.
The quarks and anti-quarks have opposite 
charge when interacting with the electric field so the field tends to
loosen the binding in mesonic bound states and, if it is strong enough, to
disassociate the mesons.
With an electric field present in the theory a \emph{singular shell}
develops in the gravity description.  If the D7 brane
passes through this shell its action becomes imaginary and 
in order to keep the action real one must turn on the appropriate electric current ($J$). 
The singular shell plays the role of an effective horizon for world volume meson fluctuations~\cite{KST}. 
If the probe brane touches the singular shell one expects that fluctuations of the
brane there must be in-falling and the spectrum will resemble a
quasi-normal mode spectrum describing mesons that have a complex mass. 
In other words the electric field has
acted to disassociate the meson in a similar fashion to how
temperature melts the mesons \cite{Peeters:2006iu}. 

The field theory phase structure is determined holographically by comparison 
of the classical bulk field configurations.  
In this paper, we introduce three bulk fields $L(\rho)$ (embedding scalar), 
$A_t(\rho)$ (gauge field), and $A_x(\rho)$ (gauge field) in the D7 brane world volume with given background parameters ($E,B,T$). 
By fixing the asymptotic values of fields in the UV (large $\rho$) 
as ($L \ra m$ (quark mass), $A_t \ra \mu$ (chemical potential), $A_x \ra 0$),  
we look for the sub-leading behaviors of fields at large $\rho$: $c$ (condensate), $d$ (density), and $J$ (current), which are 
determined by the bulk DBI dynamics. 
Therefore, our problem is classifying the phases by three quantities $(c,d,J)$ in the
5D space ($T,B,E,m,\mu$).
It turns out that, because of a scaling symmetry, we can scale all variables by $B$,
which reduces our phase space to 4D. 
Since we are interested in spontaneous chiral symmetry breaking, we will choose $m=0$.
Our phase space becomes 3D ($T,\mu,E$) and we will classify this space by 8 
possible states consisting of the three order parameters ($c,d,J$) being
``on or off''. Among them, only in the $c\ne0, d=J=0$ phase are
stable mesons allowed\footnote{We should caution that when studying phase diagrams one is always
limited by the states allowed in the analysis. We do not study the
effects of the parameters on the squark potential for example
which is likely unstable in the presence of a chemical potential 
\cite{Apreda:2005yz}.
The electric field could also potentially generate phenomena
beyond chiral symmetry restoration, density creation, current
induction, and meson melting but our results should stand as a starting point for
exploring such extra phases should they exist.}.
$T,\mu,E$ tend to turn off $c$ ($c=0$) and turn on $d,J$ ($d,J \ne 0$), and so oppose $B$. 
Due to these competitions between $E,T,\mu$ and $B$ a rich phase structure is constructed.

One of our main results here is that three phases (corresponding to three embedding types) 
are also present in the $T,E$ plane
as we summarize in Fig \ref{TvsE}.  
With a large (small) electric field, the theory becomes a
conductor (insulator) and is in $\chi$S ($\chi$SB) phase. 
At an intermediate electric field, the system is a 
chiral symmetry broken conductor.
Note that the orders of the phase transitions and the positions and presence of
the critical points vary relative to the $T,\mu$ case (Fig \ref{Tvsmu}).

Interestingly, the $\mu$-$E$ phase diagram (Fig \ref{muvsE}) shows a very different 
structure from the $T$-$\mu$ or the $T$-$E$ phase diagram.
At zero temperature and finite $E$ and $B$ field, the contribution to
the action from the density is
canceled by the contribution from the induced current, which is a function of density.  
Consequently the free energy of 
the system is independent of density, which was pointed 
out in \cite{Erdmenger1} in the zero $B$ field case.
Thus the system is essentially a zero density system and 
the phase diagram is independent of the chemical potential.                 
At first this may seem at odds with Fig \ref{Tvsmu} 
since the $\mu$ axis has structure present.
In fact there is a first order transition at $T=0$ between the $\mu$ axis
and the rest of the $E,\mu$ plane.  
There are two limits approaching the $\mu$-axis: (1) $E=0$ and then $T \ra 0$ 
(2) $T = 0$ and then $E \ra 0$. These two limits are different and only
the former is a continuous limit to the $\mu$-axis.

Indeed by computing the phase structure through 
the entire $T,\mu ,E$ volume (e.g. computing the $\mu$-$E$ diagram at various fixed $T$) 
we show the smooth evolution through that volume connecting the three surface
planes. There are interesting movements of the critical points and phase
boundaries. These results are summarized in Fig \ref{3D} which we discuss 
in much more detail in section \ref{sec6}.
We also see the development of the first order transition between 
the $E$-$\mu$ plane and the $\mu$-axis as $T \ra 0$. 
Two of the missing states from Fig \ref{Tvsmu}-\ref{muvsE}, ($c\ne0, d\ne0, J\ne0$) 
and ($c=0,d\ne0,J\ne0$) from among the 8 possible states, 
are found in the 3D bulk of Fig \ref{3D}. 

These models with their varied behaviours in the $B,T,\mu,E$
volume can hopefully serve as exemplars, or templates from which
to find exemplars, for different phase structures in physical
theories.

\section{The holographic description}

The ${\cal N}$=4 gauge theory at finite temperature has a
holographic description in terms of an AdS$_5$ black hole geometry
(with $N$ D3 branes at its
core)\cite{Malda,Witten:1998qj,Gubser:1998bc}. The geometry is
\begin{eqnarray}
  ds^2 = \frac{r^2}{R^2}(-f dt^2 + d\vec{x}^2) + \frac{R^2}{r^2 f} dr^2
  + R^2 d\Omega_5^2 \ ,
\end{eqnarray}
where $R^4=4 \pi g_s N \alpha^{'2}$ and
\begin{eqnarray}
  f := 1-\frac{r_H^4}{r^4} \ , \qquad r_H := \pi R^2 T_{\mathrm{FT}} \ .
\end{eqnarray}
Here $r_H$ is the position of the black hole horizon which is linearly 
related to the dual field theory temperature $T_{\mathrm{FT}}$.

We will find it useful to make the coordinate transformation
\begin{eqnarray}
  \frac{dr^2}{r^2f} \equiv \frac{dw^2}{w^2}
   \ \lra \  \sqrt{2} w := \sqrt{r^2 + \sqrt{r^4 - r_H^4}}\ , \label{rtow}
\end{eqnarray}
with $\sqrt{2} w_H = r_H$. This change makes the presence of a flat 6-plane
perpendicular to the horizon manifest. We will then write the
coordinates in that plane as $\rho$ and $L$ according to
\begin{eqnarray}
  w = \sqrt{\rho^2 + L^2}\ ,  \quad \rho := w\sin\theta \ .
  \quad L := w\cos\theta \ ,
\end{eqnarray}
The metric is then
\begin{eqnarray}
   ds^2 = \frac{w^2}{R^2}(- g_t dt^2 + g_x d\vec{x}^2) 
         + \frac{R^2}{w^2} (d\rho^2 + \rho^2 d\Omega_3^2
         + dL^2 + L^2 d\Omega_1^2) \ ,
\end{eqnarray}
where
\begin{eqnarray}
g_t := \frac{(w^4 - w_H^4)^2}{ w^4 (w^4+w_H^4)}\ ,  \qquad
g_x  := \frac{w^4 + w_H^4}{  w^4} \ .
\end{eqnarray}

Quenched ($N_f \ll N$) ${\cal N}$=2 quark superfields can be
included in the ${\cal N}$=4 gauge theory through probe D7 branes
in the geometry\cite{Karch,Polchinski,Bertolini:2001qa,Mateos}.
The D3-D7 strings are the quarks. D7-D7 strings holographically
describe mesonic operators and their sources. The D7 probe can be
described by its DBI action
\beq S_{DBI} = - T_{D7} \int d^8\xi \sqrt{- {\rm det} (P[G]_{ab} +
2 \pi \alpha' F_{ab})} \ , \eeq
where $P[G]_{ab}$ is the pullback of the metric and $F_{ab}$ is
the gauge field living on the D7 world volume. We will use
$F_{ab}$ to introduce a constant magnetic field (eg $F_{12} = -
F_{21} = B/(2\pi\a')$)~\cite{Johnson1}, a chemical potential associated with
baryon number $A_t(\rho) \neq 0$~\cite{Myers1,Kim} and our crucial
extra ingredient here an electric field parallel to the magnetic
field ($F_{03}=-F_{30}=E/(2\pi\a')$)\cite{Andy1, Hall, Erdmenger1, Veselin1}. We will also allow for the possibility
that the electric field induces a current in the $z$-direction by
including $A_z$.

We embed the D7 brane in the $t$, $\vec{x}$, $\rho$ and $ \Omega_3$ directions of
the metric but to allow all possible embeddings must include a
profile $L(\rho)$ at constant $\Omega_1$. The full DBI action we
will consider is then
\begin{eqnarray}
  S = \int d\xi^8 \call(\rho)
    = \left(\int_{S^3} \e_3 \int dtd\vec{x} \right) \int d\rho \
  \call(\rho) \ ,
\end{eqnarray}
where $\e_3$ is a volume form on the 3-sphere. Here
\begin{eqnarray}
  \call &:=& - \caln \rho^3\left(1-\frac{w_H^4}{w^4}\right)
   \sqrt{\left(\left(1+\frac{w_H^4}{w^4}\right)^2 + \frac{ R^4}{w^4}B^2 \right)}
  \label{OriginalAction} \\
  &&\times \sqrt{\left(\left(1-\frac{E^2R^4w^4}{(w^4-w_H^4)^2}\right)(1+ L'^2)
   - \frac{  w^4 (w^4+w_H^4)}{(w^4 - w_H^4)^2} (2\pi\a' A_t')^2  + \frac{w^4 }{w^4+w_H^4}( 2 \pi\a'  A_z')^2 \right)}
    \ , \nn 
\end{eqnarray}
and $\caln := N_f T_{D7}$. At large $\rho$, for fixed $E$ and $B$, the
fields behave as
\begin{equation}
L \sim  m + {c \over \rho^2}+..., \hspace{0.5cm} A_t \sim \mu + {
d \over \rho^2}+..., \hspace{0.5cm} A_z  \sim  {J \over \rho^2} \ ,
\end{equation}
where $m$ is the quark mass, $c$ the quark
condensate, $\mu$ the chemical potential, $d$ the quark density
and $J$ the current.
Since the action is independent of $A_t$ and $A_z$, there are
conserved quantities $d$ $\left(:= \frac{\delta S}{\delta F_{\rho
t}}\right)$ and $J$ $\left(:= \frac{\delta S}{\delta F_{\rho
z}}\right)$.  These relations can be inverted to express $A_t'$ and $A_z'$ in terms of $d$ and $J$ as  
\begin{eqnarray}
  && 2\pi \a' A_t' = \frac{d}{2\pi\alpha' \caln}\frac{ w^4 - w_H^4}{ w^4+w_H^4} Q 
  \ , \qquad 2\pi \a' A_z' = \frac{J}{2\pi\alpha' \caln}\frac{ w^4 + w_H^4}{ w^4-w_H^4} Q \  , \label{Atp} \\
  && Q := \sqrt{\frac{\left(1- E^2 R^4 \frac{w^4}{(w^4 - w_H^4)^2}\right)(1+L'^2)}{ 
  \left(\frac{d}{2\pi\alpha' \caln}\right)^2 \frac{w^4}{(w^4 + w_H^4)} 
  -\left(\frac{J}{2\pi\a' \caln}\right)^2 \frac{w^4 (w^4 + w_H^4)}{ (w^4 - w_H^4)^2 }
  + \left(\frac{B^2 R^4}{w^4} + \left(1+\frac{w_H^4}{w^4}\right)^2 \right)\rho^6   }} \ . \nn
\end{eqnarray}
This is used to express the Legendre transformed action in terms of $d$ and $J$:  
\begin{eqnarray}
  S_{LT} &=& S - \int d\xi^8 F_{\rho t} \frac{\delta S}{\delta F_{\rho t}}
  - \int d\xi^8 F_{\rho z} \frac{\delta S}{\delta F_{\rho z}} \nn \\
     &=&   \left(\int_{S^3} \e_3 \int dtd\vec{x} \right) \int d\rho \
  {\call_{LT}}(\rho) \ ,  \label{LegendreAction}
\end{eqnarray}
where
\begin{eqnarray}
   \call_{LT} &:=& - \caln \frac{(w^4-w_H^4)}{ w^4}
  \sqrt{K (1+L'^2)} \ , \label{LT} \\
      K  &:=&  \left(1-\frac{E^2R^4w^4}{(w^4-w_H^4)^2}\right) \left[\left(\frac{w^4+w_H^4}{w^4}\right)^2 \rho^6
    + \frac{  R^4 B^2}{w^4} \rho^6  \right.   \nn \\
    && \left. \hspace{1cm}+ \frac{ w^{4}}{(w^4+w_H^4)}
    \frac{d^2}{(N_f T_{D7} 2\pi\a')^2} - \frac{  w^4 (w^4+w_H^4)}{(w^4 - w_H^4)^2}  \frac{J^2}{(N_f T_{D7} 2\pi\a')^2} \right] \ .
\end{eqnarray}

Note that the first factor of $K$ changes sign at $w_s$,
\begin{eqnarray}
  w_s = \sqrt{\frac{E R^2}{2} + \frac{\sqrt{E^2 R^4 + 4 w_H^4}}{2}} \ , \label{ws}
\end{eqnarray}
which defines a \emph{singular shell} with a radius $w_s$.
At zero temperature ($w_H=0$) the singular shell forms at $ \sqrt{E} R $,
and at zero $E$, it merges into the horizon $w_s = w_H$.
Note that the singular shell does not depend on density.
In order to make the action regular, the second term of $K$ should change
sign at the singular shell. This condition determines the current $J$ and conductivity
$\sigma$ by Ohm's law:
\begin{eqnarray}
  && J = \sigma E \ , \label{Ohm} \\
  && \sigma := \caln (2\pi\a') R^2 \sqrt{\frac{ w_s^4}{(w_s^4 + w_H^4)^2} \frac{d^2}{N^2 (2\pi\a')^2}
  + \left[\frac{ R^4}{w_s^4 (w_s^4 + w_H^4)} B^2
  + \frac{ (w_s^4 + w_H^4) }{w_s^8} \right] \rho_s^6} \ , \nn
\end{eqnarray}
where $\rho_s$ is the $\rho$ coordinate where an embedding touches
the singular shell. This is still a function of quark mass $m$
after all other parameters are fixed. Inspite of the way we write
(\ref{Ohm}) the current is non linear in $E$ since $w_s$ and
$\rho_s$ are functions of $E$. $\sigma$ has two contributions. The
first term is from net charge carrier density, $d$, and the second
term is from pair produced virtual charges. Interestingly, the
conductivity by pair produced charges are enhanced by $B$. 
The more general conductivity for arbitrarily angled constant $E$ and $B$ 
was obtained in \cite{Hall,Andy3} in a different coordinate system\footnote{The conductivity 
of other models have been obtained by the same method. See for example \cite{KE} and references therein. }.
By plugging (\ref{Ohm}) into (\ref{LT}) and rescaling we have a dimensionless Lagrangian
$\wt{\call}_{LT}$:
\begin{eqnarray}
   && \wt{\call}_{LT} := -\frac{\call_{LT}}{R^4 B^2\caln} =  \frac{(w^4-T^4)}{ w^4}
  \sqrt{\wt{K} (1+L'^2)} \ ,\label{Ham} \\
    &&  \wt{K}  :=  \left(1-\frac{E^2 w^4}{(w^4-T^4)^2}\right) \left[\left(\frac{w^4+T^4}{w^4}\right)^2 \rho^6
    + \frac{1}{w^4} \rho^6  +  d^2 \frac{ w^{4}}{(w^4+T^4)} \right. \nn \\
    && \left. \hspace{1cm}
       - E^2 \frac{  w^4 (w^4+T^4) (w_s^4 + (w_s^4+T^4)^2 )}{(w^4 - T^4)^2(w_s^4+T^4) w_s^8 } \rho_s^6
      - d^2 E^2 \frac{w^4(w^4+T^4)w_s^4}{(w^4-T^4)^2(T^4+w_s^4)^2} \right] \ ,
\end{eqnarray}
where we rescaled
\begin{eqnarray} \label{rescale}
  (\w, L, \rho) \ra R \sqrt{B}\ (\w, L, \rho) \ ,  \quad 
  (d, J) \ra (R \sqrt{B})^3 \caln 2\pi \a' \ (d, J) \ ,  \quad  E \ra BE \ ,
\end{eqnarray}
and define $T \equiv w_H$ for notational clarity.  
The Lagrangian $\wt{\call}_{LT}$ will be our
starting point for the numerical analysis in the following
sections.

The chemical potential is obtained by integrating $A_t'$ (\ref{Atp})
from the horizon to the boundary
\begin{equation}
\mu = \int_{\rho_H}^{\infty} d \frac{ w^4 - T^4}{ w^4+T^4} 
\sqrt{\frac{\left(1- E^2 \frac{w^4}{(w^4 - T^4)^2}\right)(1+L'^2)}{ 
  d^2 \frac{w^4}{(w^4 + T^4)} 
  -J^2 \frac{w^4 (w^4 + T^4)}{ (w^4 - T^4)^2 }
  + \left(\frac{1 }{w^4} + \left(1+\frac{T^4}{w^4}\right)^2 \right)\rho^6   }} \ , \label{mu0}
\end{equation}
where $A_t(\rho_H) = 0$.

In the following sections we will present our results on various 
aspects of the phase structure of this theory. Until the final section we will
concentrate on the case of massless quarks where the $U(1)$ symmetry
in the $d\Omega_1$ direction is a good UV symmetry of the theory.
Also here and below we will express all our dimensionful
parameters in units of the magnetic field $B$ to the appropriate
power (see (\ref{rescale})) - in other words we will use the
magnetic field as the intrinsic scale in the theory.

\section{$B$, $T$, $\mu$ Phase Diagram}

In this section we review previous results \cite{Evans1} on the theory without
an electric field present. The influence of $B$, $\mu$ and $T$ on the D7 embedding can be
understood qualitatively as follows. With none of these terms
present the embedding of the ${\cal N}=2$ theory for massless
quarks is just flat, $L(\rho)=0$ - $c$ is zero and there is no
quark condensate generated to break the $U(1)$ symmetry.

Temperature is represented by an attractive black hole at small
$\rho$ and also favours a flat embedding.
The presence of density $d$ leads to solutions that spike to the
origin (or on to the black hole horizon with temperature present)
- the spike can be thought of as strings linking the D3 and D7
branes and represent the density of quarks explicitly. Note that
at finite chemical potential there can also be solutions with
$A_t=\mu$ and $d=0$ - the action for such solutions is independent
of $\mu$ and hence so are the embeddings.

Non-trivial embeddings are generated by the $B$ field. We can see
from (\ref{OriginalAction}) that the last square root term grows
if the embedding approaches $w=0$ when there is a $B$ field. The
embeddings tend to curve off axis to avoid the origin leading to a
non-zero quark condensate $c$ and spontaneous breaking of the $U(1)$
symmetry.
\begin{figure}[]
\centering
\subfigure[The D7 brane embeddings]
  {\includegraphics[width=4.9cm]{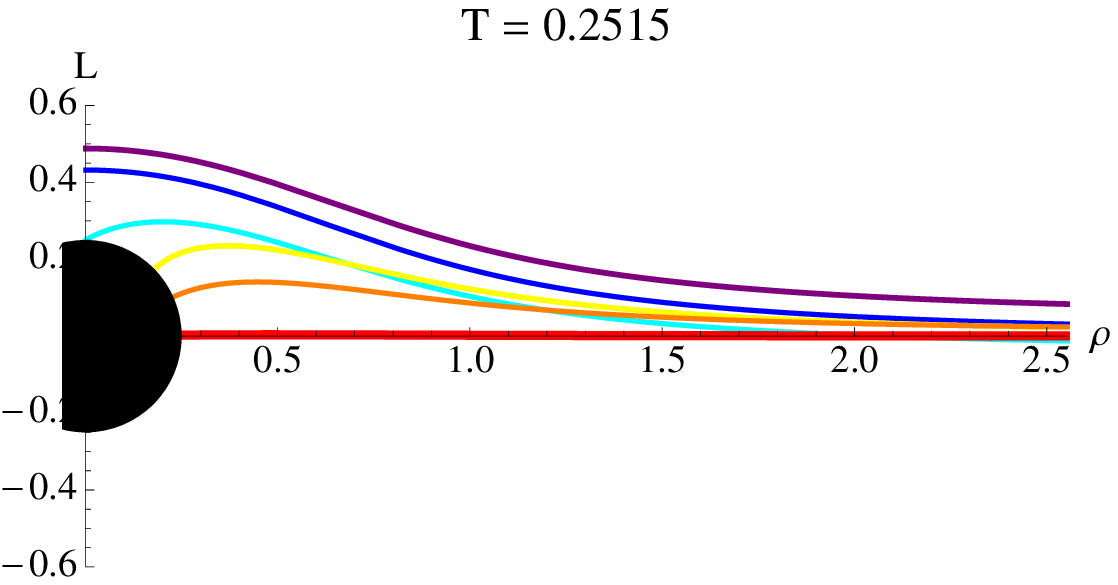}\label{Fig2a}}
\subfigure[$m$- $c$ diagrams ]  
   {\includegraphics[width=4.9cm]{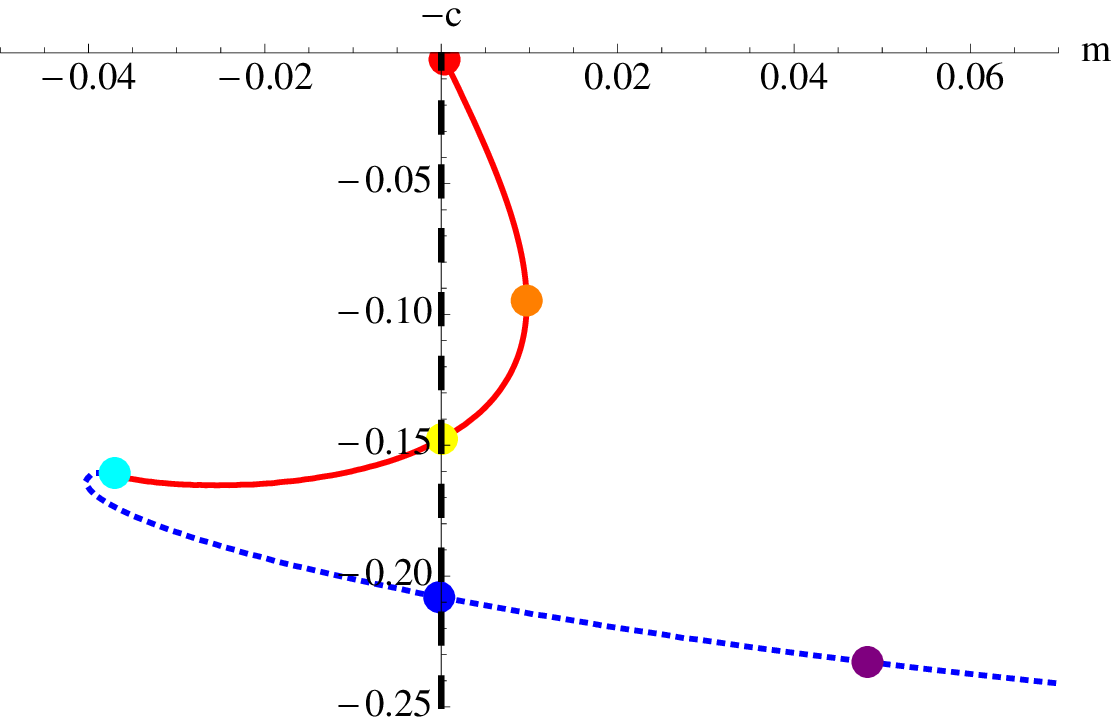}\label{Fig2b}}
\subfigure[the Free energies]   
   {\includegraphics[width=4.9cm]{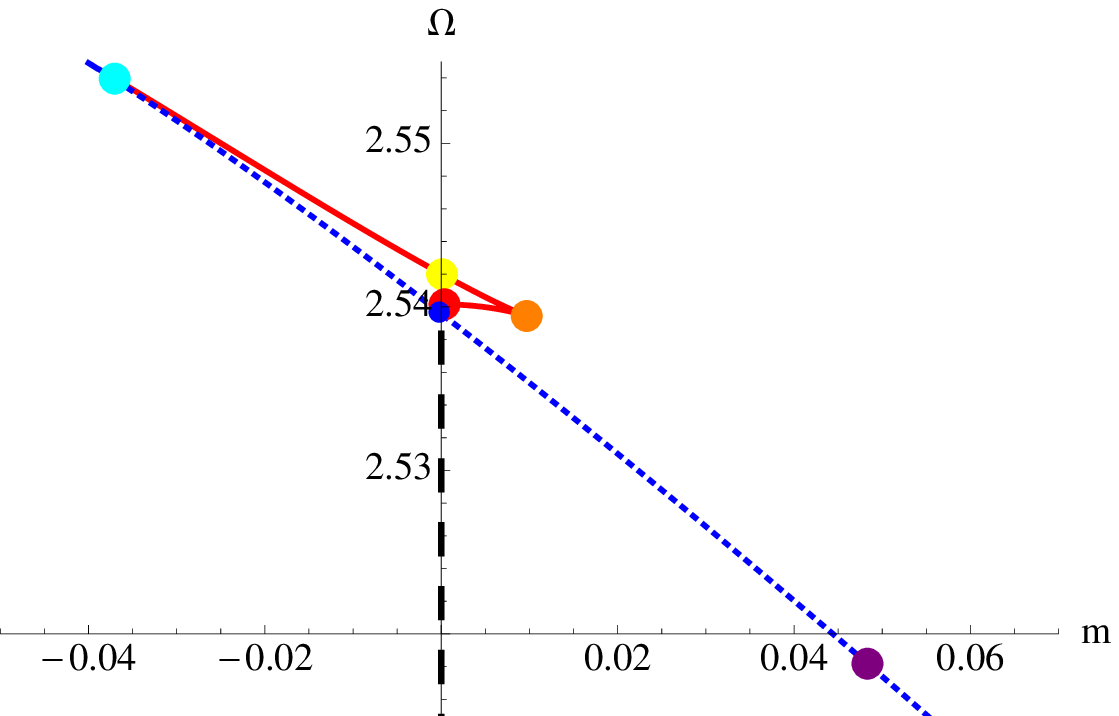}\label{Fig2c}}
  \caption{ Exemplary plots in the presence of a magnetic field at finite chemical potential and with
            temperature $T=0.2156$. (Parameters are scaled by $B$.)
           }\label{Fig2}
\end{figure}

The interesting phase structure of the theory is generated by the
competition between the $B$ field wanting to break the $U(1)$ symmetry
and $d$ and $T$ which seek to restore it. As an example we show in
Fig \ref{Fig2a} solutions of the Euler-lagrange equations for the
embedding for a case with $\sqrt{B} > T
> \mu$ - the red embedding is flat and preserves the symmetry, the
yellow embedding spikes on to the horizon and the dark blue
embedding breaks the $U(1)$ symmetry and has $d=0$. The full phase
diagram is generated by competition between these three types of
embeddings.

To decide which of the embeddings represents the true vacuum one
may compute the free energy $\Omega = - S$ evaluated on the
solutions. This plot is shown on the right hand side of Fig \ref{Fig2c}.
Here we see that we have chosen a transition point where the red
and dark blue curves are degenerate - for larger $T$ the flat
embedding is preferred. Here this transition is associated with
chiral symmetry restoration ($c$ changes from non-zero to zero)
and melting of the mesonic degrees of freedom (the brane moves
from off the horizon to on it). It is a first order transition.

The transition's presence and position in parameter space can also
be seen from the middle plot in Fig \ref{Fig2b}. Here we plot the
condensate $c$ against the quark mass $m$. To obtain this plot we
must find solutions for the embeddings that lie off axis as $\rho
\rightarrow \infty$ such as the green and orange curves shown. We
can see the presence of the three solutions discussed at $m=0$ in
the middle plot. To see there is a phase transition at this point
we can use a Maxwell construction. The quark mass and condensate
are conjugate variables and therefore the area between the $c$
axis and the curve represents the free energy difference between two
points on the curve. At this point the two segments of the curve
generate equal area and the first order transition between the red
and blue points is predicted.

In \cite{Evans1} we fleshed out in this way the full phase diagram of the
theory which we show in Fig \ref{Tvsmu}. We quickly present the
results here since the analysis is a subset of the work in section
\ref{sec6} below. In the bottom left segment the symmetry breaking, black
hole avoiding embedding is the vacuum; in the central section a
spike embedding is preferred; and to the right the symmetry
preserving flat embedding is the vacuum. The low $T, \mu$ phase
has chiral symmetry breaking, no density and stable meson states; 
the central regime chiral symmetry breaking, finite density and melted mesons; 
in the high $T,\mu$ phase the symmetry is restored and the mesons melted at finite density. 
The order of the transitions is marked as well and each transition has
periods where it is first and second order, linked by
critical points.

\section{$B$, $T$, $E$ Phase Diagram}

The rich structure of the $B, \mu, T$ phase diagram leads one to
ask how generic it is? The main goal of this paper is to introduce
an additional parameter that favours chiral symmetry preservation
to see how sensitive the phase diagram is to a change of
parameter. We will use electric field ($E$) as that new parameter.

\subsection{$B$, $E$ at zero temperature} \label{S1}

As a first example lets consider the system with $E$ and $B$ but
no $T$ or $\mu$. The Legendre transformed Lagrangian is
\begin{equation}
   \wt{\call}_{LT} =
  \sqrt{(1+L'^2)}\sqrt{ \left(1-\frac{E^2}{w^4}\right) \left[ \rho^6
    + \frac{\rho^6}{w^4}  -  J^2 \right]}  \ . \label{BE}
\end{equation}
As has been discussed in (\ref{ws}) and (\ref{Ohm}), there is a
singular shell at $w_s = \sqrt{ E }$ and the current is given by
\begin{equation}
J= E \frac{\sqrt{1+w_s^4} \rho_s^3}{w_s^4}  =  \sqrt{(1+E^2) E \cos^6 \theta_s} \ . \label{BEJ}
\end{equation}
\begin{figure}[]
\centering
\subfigure[Embeddings ending on and off the singular shell ($E=0.1, m=0$).]
  {\includegraphics[width=11cm]{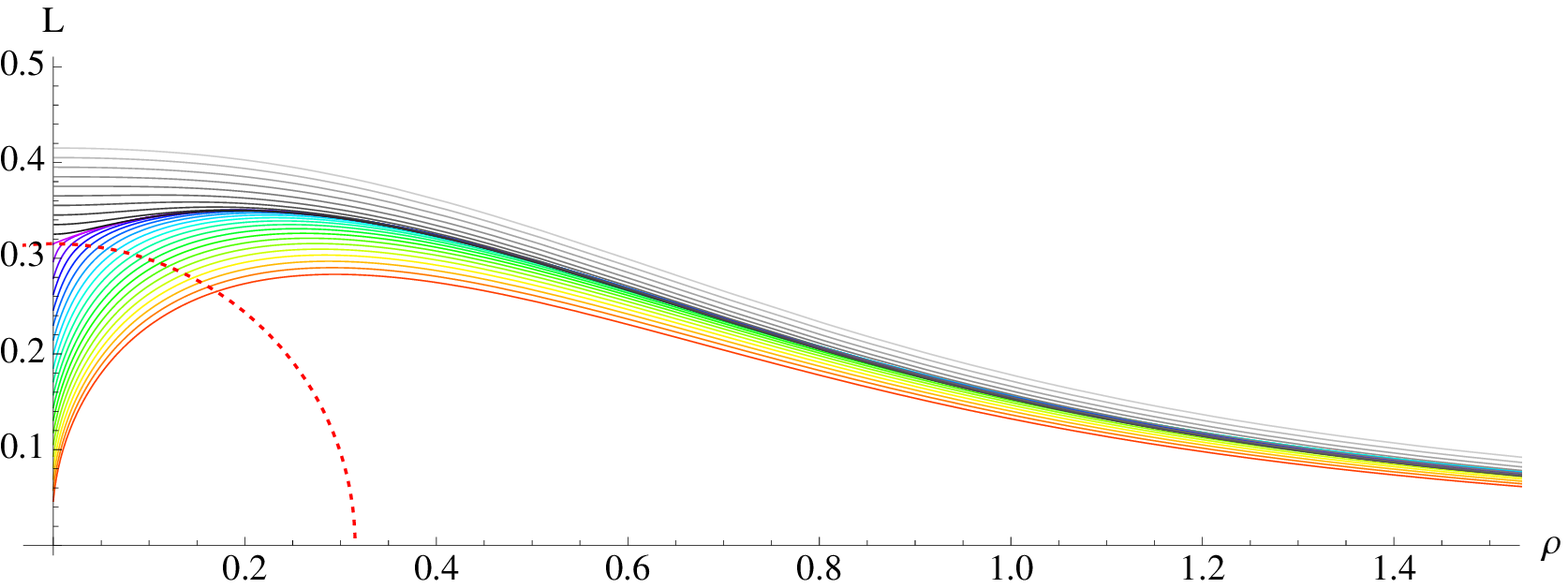}\label{BE.embedding}}
\subfigure[The condensate, $c$ vs mass, $m$ for given $E$ values. 
Curves shift to left as $E$ decreases. 
The red (blue) part corresponds to the embedding touching (missing) the singular shell.    
The inset on the right is a zoom-in around the first order phase transition. 
The colored points correspond to the colored embeddings in Fig \ref{BE.embedding}. ]  
   {\includegraphics[width=7cm]{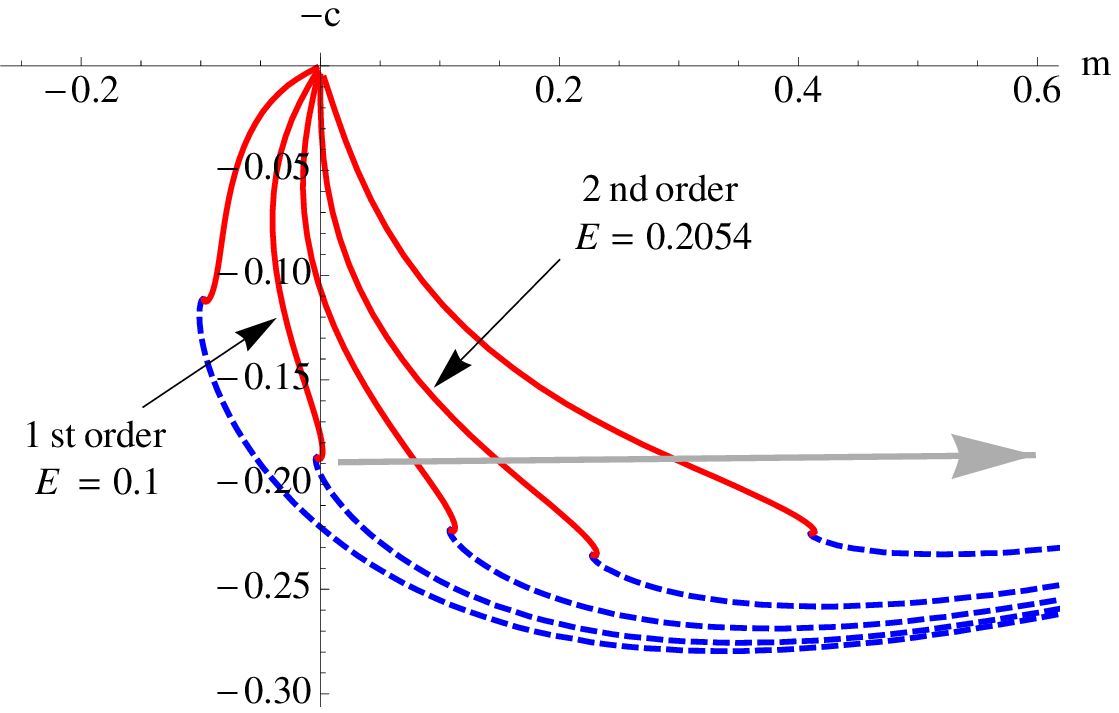}
    \includegraphics[width=6cm]{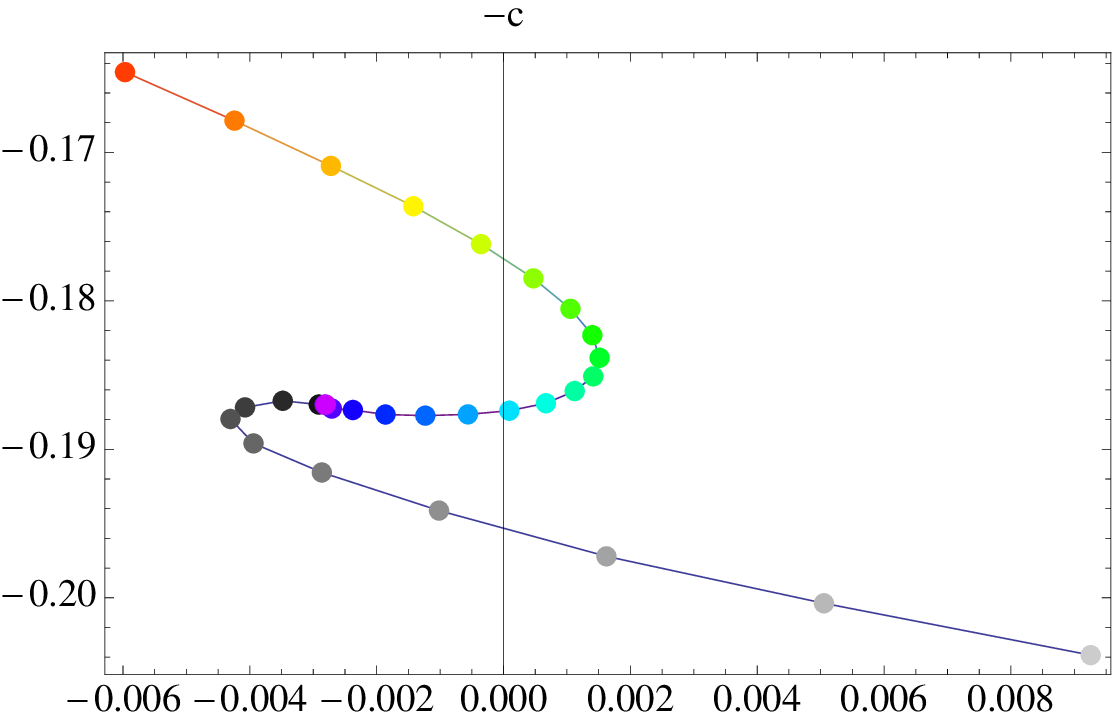}\label{BE.mc}}
  \caption{{ Embeddings and $c$-$m$ diagram at finite $E$ parallel to finite $B$, but no density and temperature.}
           }%\label{Fig.BE}
\end{figure}

One can therefore find the embeddings that end on the singular
shell by shooting out (in) from the singular shell with $J$
determined by the point on the shell one is shooting from. 
We then seek amongst such solutions for those that satisfy $L \rightarrow
0$ ($m \ra 0$) as $\rho \rightarrow \infty$ to find the massless (mass
$m$) embeddings. Generically we again find three types of
solutions - (1) embeddings that curve off axis and reach $\rho=0$
missing the singular shell (when $E \ll B$); (2) embeddings that curve
off axis and pass through the singular shell (when $E \sim B$); (3) the flat
embedding $L=0$. The schematic plots of 
these three cases are shown in the inset of Fig \ref{TvsE}, where
the black disk should be ignored at zero $T$. 
In Fig \ref{BE.embedding} we show some sample numerical embeddings 
ending on (2) and off (1) the singular shell.

All non-flat embeddings that pass through the singular shell
have a conical singularity at $\rho = 0$, whose precise interpretation 
is unclear and discussed in~\cite{Erdmenger1, Veselin1}.
The conical singularity is most likely a reflection of the 
energy being injected by the electric field being sunk into the 
gauge background through stringy physics representing the quark interactions
with the ${\cal N}=4$ Yang-Mills fields.

The Fig \ref{BE.mc} shows the condensate, $c$, vs mass, $m$,
plot for given $E$s. One point ($c$,$m$) in the plot 
corresponds to one embedding since it gives a complete 
initial condition for the embedding equation 
(a second order differential equation).
For example, the colored points in the inset of Fig \ref{BE.mc} correspond 
to the embedding in Fig \ref{BE.embedding} of the same color. 

From Fig \ref{BE.mc} we can determine the phase structure as follows. 

For a larger $E$ (relative to $B$), the $c$-$m$ curves tend to be  
pushed to the right. So if we focus on $m=0$ case, the only 
available point at large $E$ is $(m,c) = (0,0)$, 
which corresponds to the flat embedding.  
The flat embedding preserves the $U(1)$ chiral symmetry, 
so the system is chiral symmetric ($\chi$S). Since the flat embedding necessarily crosses the singular shell,
there is no stable meson but there is a current (\ref{BEJ}) with $\theta_s = 0$: $J = \sqrt{(1+E^2)E}$, which is
the maximal current for a given $E$. i.e. The system is a conductor. 

As $E$ is lowered the $c$-$m$ curve shifts to the left. 
At $E\sim0.2054$ a new solution for $m=0$ appears, whose 
$c$ is finite and the ground state corresponds to this new solution. 
This is a second order transition  because the second solution for $c$ 
at $m=0$ moves smoothly away from the $c=0$ embedding.
Because of the finite $c$, the embedding is curved and breaks the $U(1)$ symmetry, 
so the system is in a chiral symmetry broken ($\chi$SB) phase. However the embedding 
still passes through the singular shell, so it is a conductor with no stable meson. 

Finally, around $E \sim 0.1$ the small `S' shape structure connecting the
red part (singular shell touching embedding) and blue part 
(a singular shell missing Minkowski embedding), 
meets the $m=0$ line, which is 
zoomed in, in the inset of Fig \ref{BE.mc}. It shows 
a typical first order phase transition structure and 
by a Maxwell construction we can pin down the 
transition point. 
So as $E$ decreases, the green point should move to the gray point 
with a discontinuous condensate jump (see the zoomed in inset). 
The blue part (or the gray points in the inset) 
corresponds to the Minkowski embedding and the system 
is a $\chi$SB insulator with stable mesons.   

One would like to match this picture to a computation of the free
energy. Naively it seems one should just compute the original
action (before Legendre transforming) evaluated on these
solutions. 
However, there is a subtle point related to the 
conical singularity\footnote{If we consider the
finite density system this conical singularity disappears. 
However another singularity seems to appear for $A_t$ as discussed 
in section \ref{sec5}.} of the embedding at $\rho=0$ 
and also a $\log$ divergence of $A_x$ \cite{Andy1, Andy2, Andy3} at the horizon. 
Thus we need to add some boundary term 
to take care of the singularity at $\rho=0$ or at the horizon.  
This boundary term will also contribute to the free energy so must be taken into account.

However, this surface term does not change the equation of motion. 
Furthermore, as far as the embedding dynamics is concerned, 
the singular shell position has the same singular structure as a black hole horizon and the 
embedding outside a singular shell is independent of the ones inside the shell.
So the $c$-$m$ plots, based on the classical embedding outside the shell, are valid regardless of 
the additional boundary terms at the IR boundary. 

These solutions correctly show us the maxima
and minima of the free energy as a function of $E$. The discussion
above is the only consistent picture with the $c$-$m$ plots so
we can be confident of its validity. For this reason in what
follows we will focus on the $c$-$m$ plots (and also when
chemical potential is present we will track the  quark density)
to determine the phase structure. A similar philosophy was used in
~\cite{Erdmenger1, Veselin1}.

However, it would be interesting to identify the 
correct boundary term and compute the consistent free energy graph 
for our $c$-$m$ plot (we have not been able to so far). 
To identify it, in principle, one should start with 
the time-dependent back-reacted system, since the singularities 
are related to time-dependent energy loss of the system and 
its effect on the background adjoint matters. 
However one may also be able to introduce an ``effective'' boundary term.   
Our $c$-$m$ diagram would be a good guide  
to figure out the correct boundary term and free energy or could even  be 
used as a rule to determine it, since sometimes thermodynamic consistency
plays a complementary role in AdS/CFT applications.

\subsection{$B$, $E$ at finite temperature}

We can now extend our analysis to include temperature straight
forwardly.
From (\ref{Ham}) with $d=0$
\begin{eqnarray}
   && \wt{\call}_{LT} =  \frac{(w^4-T^4)}{ w^4}
  \sqrt{\wt{K} (1+L'^2)}  , \nn \\
   &&   \wt{K}  =  \left(1-\frac{E^2 w^4}{(w^4-T^4)^2}\right)  \left[\left(\frac{w^4+T^4}{w^4}\right)^2 \rho^6
    + \frac{1}{w^4} \rho^6
       -  J^2 \frac{  w^4 (w^4+T^4)}{(w^4 - T^4)^2}
      \right] \ ,
\end{eqnarray}
and
\begin{eqnarray}
  w_s = \sqrt{\frac{E }{2} + \frac{\sqrt{E^2  + 4 T^4}}{2}} \ ,  \qquad
  J =  E \sqrt{\frac{w_s^4 + (w_s^4+T^4)^2}{(w_s^4+T^4) w_s^8 } } \rho_s^3 \ .
\end{eqnarray}
It is apparent that even with non-zero $T$ there remains a singular shell
- it always lies outside $T$ for any $T$ (\ref{ws}). Requiring regularity of 
the onshell action allows us to fix the current $J$ (\ref{Ohm}).
If the embedding does not touch the singular shell $\rho_s = 0$, then there is
no current.

\begin{figure}[]
\centering
\subfigure[$T=0.24$. $E=0.0001, 0.049, 0.1$ from left.]
   {\includegraphics[width=7cm]{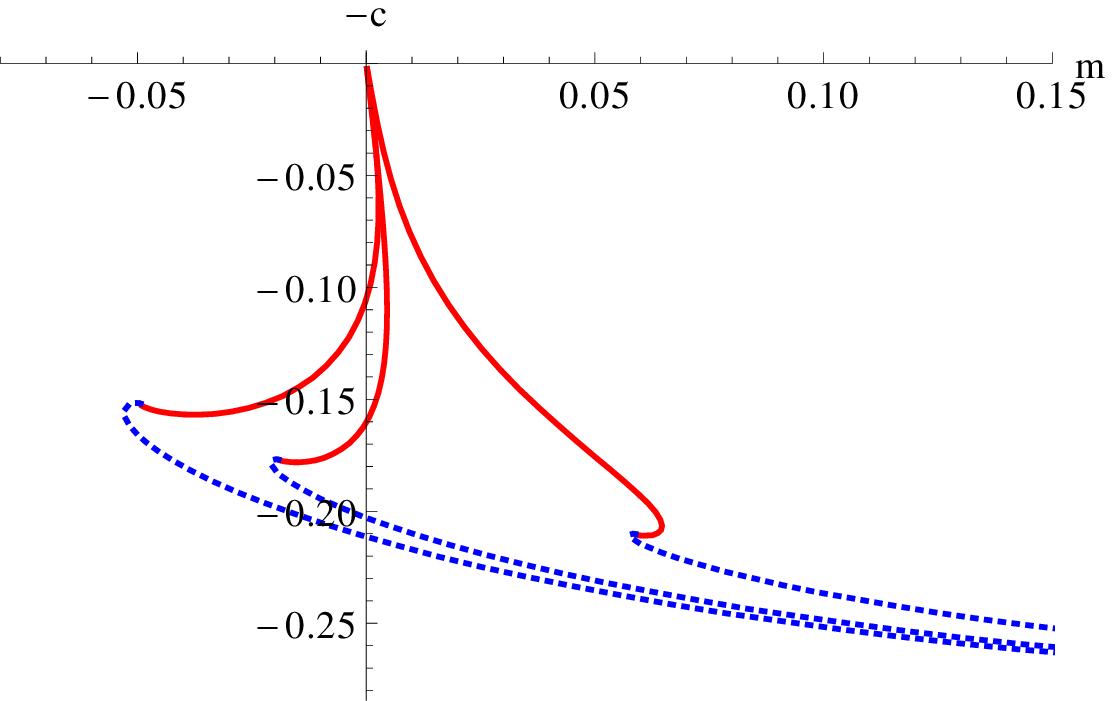}\label{cm2a}}
\subfigure[$T=0.30$. $E=0.0001, 0.100, 0.180$ from left.]   
   {\includegraphics[width=7cm]{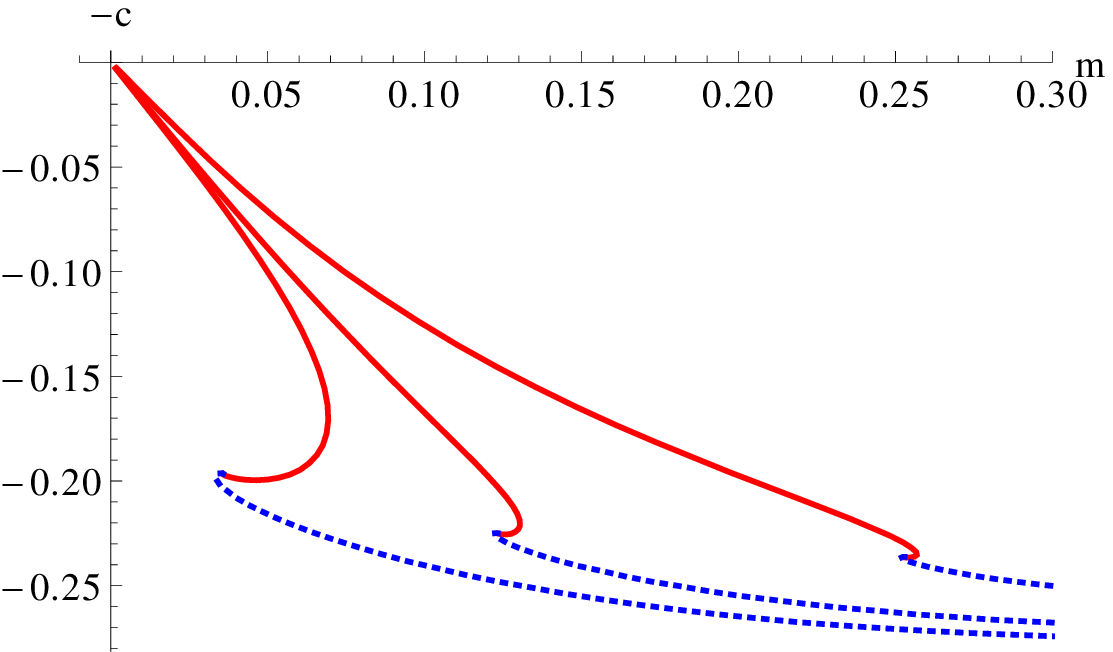}\label{cm2b}}
  \caption{{ $c$-$m$ diagram at finite $E$ parallel to finite $B$ at finite (high) temperature.}
           }\label{cm2}
\end{figure}

We shoot out to obtain the embeddings as a function of $E$ and $T$
at fixed $B$. The process is laborious - we plot the evolution of
the $c$-$m$ plot on fixed $T$ trajectories as we did at T=0 in
the previous section. 
There are three types of $c$-$m$ plot. 
At low temperature, it is similar to Fig \ref{BE.mc}. 
At high temperature two qualitatively different structures appear as shown in 
Fig \ref{cm2}. As the temperature increases, the $T$ effect dominates 
the $E$ effect, which is visualized in the $c$-$m$ diagram as follows. 
The curve near $(m,c)=(0,0)$
is curved to the left (Fig \ref{BE.mc}) at low $T$ but curved to the right (Fig \ref{cm2a}) at high $T$
(similarly to Fig \ref{Fig2}),    
showing competition between $E$ and $T$. Both are an attractive effect from the embedding dynamic's
point of view, but the $T$ driven 1st order attraction is so strong 
that the $E$ driven 2nd order smooth attractive effect cannot be realized.  
At very high $T$ the repulsive effect of $B$ is completely suppressed and 
the only allowed embedding is a flat one (Fig \ref{cm2b}).  

The resulting phase diagram is Fig \ref{TvsE}.
For $T \lesssim 0.233$, three regions, as at zero $T$, exist: as $E$ increases, the phases changes from a
$\chi$SB insulator (stable meson) phase to a
$\chi$SB conductor at a first order transition. There is then a second order transition to a $\chi$S 
conductor. 
Above $T \sim 0.233$ the intermediate region $\chi$SB and conductor phase disappears and 
the chiral symmetry restoration and insulator-conductor transition happen at the same time. It is a first order
transition and exists in the temperature range
 $0.233 \lesssim T \lesssim 0.25$.
Finally for higher temperatures $T \gtrsim 0.25$, the system becomes a chiral symmetric conductor for any
finite $E$.  

There are distinct features of the phase diagram though from the $T, \mu$ phase diagram in Fig
\ref{Tvsmu}. The insulator-conductor transition is first order
along its whole length.
At finite mass and zero magnetic field, this insulator-conductor transition was 
also shown to be first order in~\cite{Veselin1,Erdmenger1}. 
The chiral symmetry restoration phase
transition is second order along all its length from the critical
point where it joins the insulator-conductor transition. 
The ability
to reproduce different phase structures is interesting and
potentially useful if one wanted to use these models as effective
descriptions of more complex gauge theories or condensed matter
systems.

\section{$B$, $\mu$, $E$ at zero temperature}\label{sec5}

We now turn to the $E,\mu$ plane at fixed B and zero $T$ where the behaviour 
appears somewhat
different from those of the planes so far discussed.

From (\ref{Ham})  with $T=0$
\begin{equation}
   \wt{\call}_{LT} =
  \sqrt{(1+L'^2)}\sqrt{ \left(1-\frac{E^2}{w^4}\right) \left[ \rho^6
    + \frac{\rho^6}{w^4} +   d^2 - J^2 \right]} \ , 
\end{equation}
and 
\begin{eqnarray}
 J =  \frac{\sqrt{d^2 w_s^4 + (1+w_s^4) \rho_s^6} }{w_s^2} 
   =   \sqrt{d^2 + (1+E^2) E \cos^6 \theta_s} \ , \label{JJ}
\end{eqnarray}
where $w_s = \sqrt{E}$. 
Note that this formula for $J$ is only valid for finite $E$,
because the current $J$ is introduced to make a sign change when
$\left(1-\frac{E^2}{w^4}\right)$ changes sign. If $E=0$ we
wouldn't have any reason to introduce $J$.

If we now substitute $J$ back into $\wt{\call}_{LT}$ then the $d$
dependence explicitly vanishes, as also observed in~\cite{Erdmenger1} 
for the zero $B$ case.
\begin{eqnarray}
   \wt{\call}_{LT} =
  \sqrt{(1+L'^2)}
  \sqrt{ \left(1-\frac{E^2}{w^4}\right) \left[ \rho^6
    + \frac{\rho^6}{w^4}  -  (1+E^2) E \cos^6 \theta_s \right]} \ .\nn 
\end{eqnarray}
It is the same action as (\ref{BE}),
but the physics could still be different because at finite density the boundary condition 
for the embedding is different from zero density.  Furthermore the current (\ref{JJ}) looks different 
from the $d=0$ case (\ref{BEJ}) through its $d$ dependence (explicitly and implicitly through $\theta_s$). 
However, at zero temperature and finite $E$, 
it turns out that the density is always zero, which can be shown as follows.
From (\ref{Atp}) or (\ref{mu0}), at zero $T$,  
\begin{eqnarray}
  A_t' = d \sqrt{\frac{(w^4 - E^2)(1+L'^2)}{d^2 w^4 + \left[- d^2 -\left(1+\frac{1}{w_s^4}\right)\rho_s^6 \right]_{E\ne0} w^4
  + (1+w^4)\rho^6    }}  \ , \label{At00}
\end{eqnarray}
where $w = \sqrt{L^2 + \rho^2}$ and the square bracket is the current which must be zero at $E=0$. 
So, for non-zero $E$, the density cancels in the denominator. 
Let us first consider a fixed non-zero $E$ and the flat embedding and take the limit $\rho \ra 0$. 
\begin{eqnarray}
  A_t' \sim \frac{d}{E+1/E} \frac{1}{\rho^2} \ .
\end{eqnarray}
Thus $\mu \sim d$ $\times$ a $d$-independent integral that diverges near $\rho=0$. 
There is, therefore,  no way to get a finite $\mu$ from a finite $d$. 
The only available density is exactly zero and if density is zero we shouldn't use the relation (\ref{At00}). 
Any constant $A_t$ is an available solution so any constant $\mu$ is allowed.

\begin{figure}[]
\centering
  {\includegraphics[width=8cm]{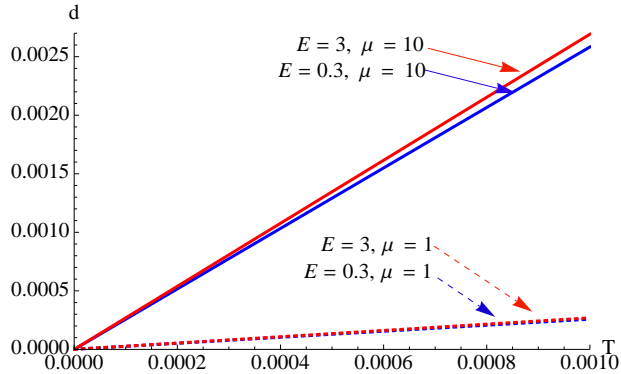}}
  \caption{ $d$ vs $T$ for four points $(E,\mu) = (3,10),(0.3,10),(3,1),(0.3,1)$. $d \ra 0$ as $T \ra 0$.
           }\label{dvsT}
\end{figure}

To confirm this analysis, we numerically evaluate density at small temperatures 
for four sample points,  $(E,\mu) = (3,10),(0.3,10),(3,1),(0.3,1)$, in Fig \ref{dvsT}.
The density indeed vanishes as $T$ goes to zero. 
Turning on larger chemical potential does not change the tendency: a
$10$ times larger chemical potential vanishes $10$ times faster (compare $\mu=10$ and $\mu=1$). 
Electric field does not affect this much: $E=3$ and $E=0.3$ at $\mu=1$ are indistinguishable in  Fig \ref{dvsT}. 
Thus we find that vanishing density at $T=0$ is consistent with the limit $T \ra 0$.

Given this argument it is worth checking how nontrivial results come from the same expression
on the $\mu$ axis in Fig \ref{Tvsmu}, where $T = E = 0$.
Let us consider $E=0$ not $E \ra 0$.  
\begin{eqnarray}
   A_t' = d \sqrt{\frac{w^4(1+L'^2)}{d^2 w^4 + (1+w^4)\rho^6    }}  \ .
\end{eqnarray}
For a flat embedding and near $\rho = 0$ 
\begin{eqnarray}
  A_t' \sim 1-\frac{\rho^2}{2 d^2} \ ,
\end{eqnarray}
We learn that $\mu \sim d$ $\times$ a $d$-dependent regular integral. So here there is a non-trivial
$\mu,d$ relation. We should be 
careful with the $d \ra 0$ limit. Then the $d$-dependent integral diverges as $d$ gets smaller. 
However it turns out that it is less divergent than $1/d$ so we get $\mu=0$ as expected. 
If we consider the spiky embedding then the singular integral will fall as $\sim 1/d$ 
due to the divergence in $L'$. Here we can have a nonzero chemical potential when $d \ra 0$. 
This occurs at the first transition point (from the solution lying outide the singular shell to the
spiky embedding) 
shown on the $\mu$ axis (zero $T$) in fig \ref{Tvsmu}.

Notice that the $T=E=0$ theory and the $T=0$, $E \rightarrow 0$ limit are distinct.
The $T=0$, $E,\mu$ plane for any finite $E$ has zero density.
However, the strict $E=0$ axis does have density present for the spike and flat
embeddings - we show this with the yellow and green lines respectively on the $E=0$ axis in Fig \ref{muvsE}. 
Since there is a jump in the density off the $E=0$ axis at $T=0$ 
there is formally a first order transition with increasing $E$, which is expressed by the blue line near 
$E=0$ in  Fig \ref{muvsE}. 
(note the transitions on the $E=0$ axis are second order though - two red dots in  Fig \ref{muvsE}). 
In the next section we will approach the $T=0$ plane from positive $T$ to confirm this picture.

The surprising aspect of this result is that, at zero $T$, and with even an infinitessimal $E$ present,
density is not generated no matter how large the chemical potential $\mu$ is. Therefore  
the flat, chirally symmetric configuration is not favoured for very large $\mu$ when a small $E$ is present. 

These conclusions are certainly correct within the DBI analysis presented here. If
the reader wishes to seek additional physics that might generate density at $T=0, E\ne0$ 
and more simply connect the phases at zero and infinitessimal $E$, then one might be able to do that through
additional boundary terms at the origin (this is where singular behaviour
also enters in the $\mu,d$ relation). Presumably
at this point the physics associated with the sink of the energy being injected by the $E$ field
should be better understood. Equally the DBI action may not be valid 
for this system due to a divergent gauge field near the origin~\cite{KST}.  
Resolving this issue is beyond our DBI analysis here.

\section{The Full $B$, $T$, $E$, $\mu$ Phase Structure} \label{sec6}

Our final task is to complete the phase structure analysis by
extending it to the full $E, \mu, T$ volume at fixed $B$. The
number of embeddings that must be analyzed on any fixed plane
through this space is already large so we restrict ourselves to
looking at some representative slices that will be sufficient to
reveal the structures present.

In particular we will study fixed $T$ slices and draw the phase
structure in the $E,\mu$ plane. The embedding equations are now
given by the full forms of (\ref{Ham}). 
To determine the presence and nature of a transition it is
sufficient to track any operator of the theory. We have found it
easiest on these planes to plot the density $d$ against the chemical 
potential $\mu$. i.e. we use $d$ as our order parameter.

\begin{figure}[]
\centering
\subfigure[2nd $\ra$ 1st order]
  {\includegraphics[width=4.5cm]{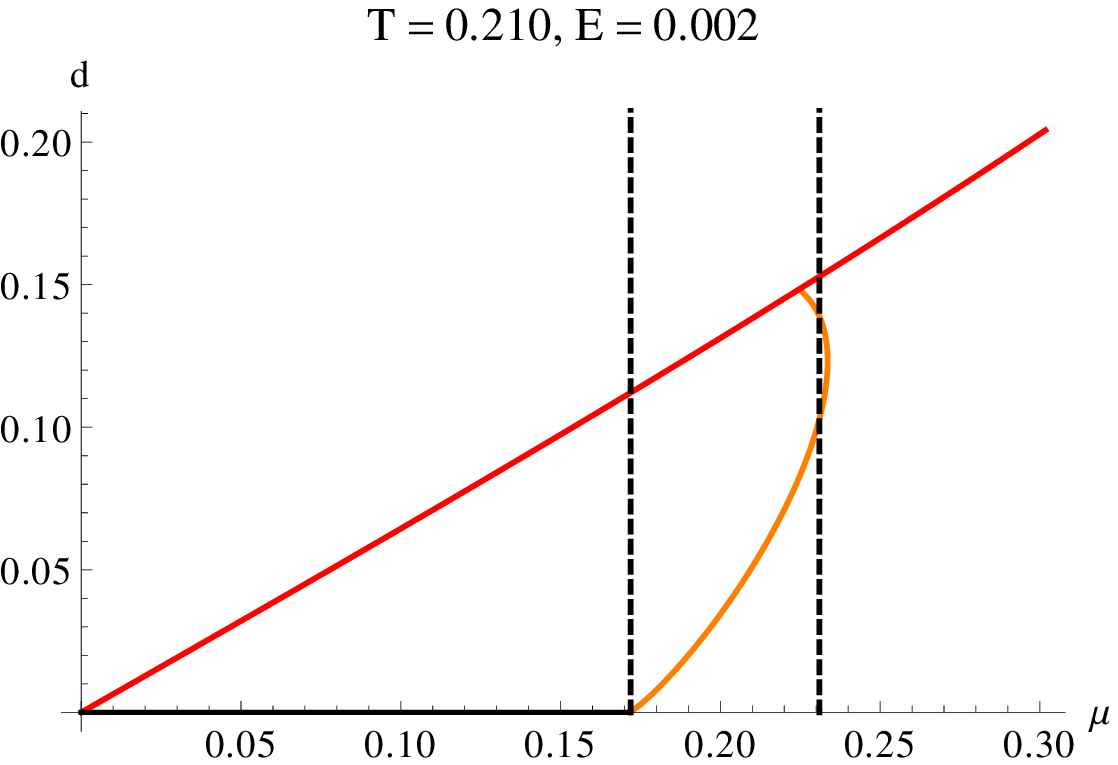}}
  \subfigure[1st $\ra$ 2nd order]
   {\includegraphics[width=4.5cm]{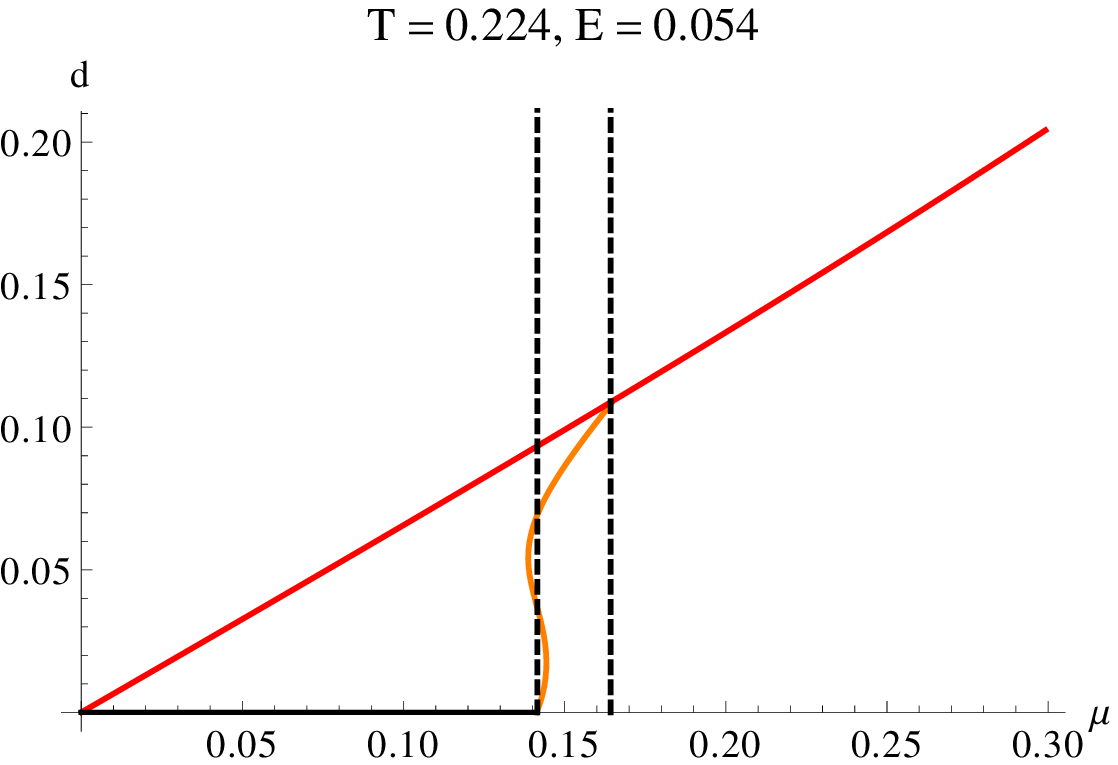}}
   \subfigure[2nd $\ra$ 2nd order]
   {\includegraphics[width=4.5cm]{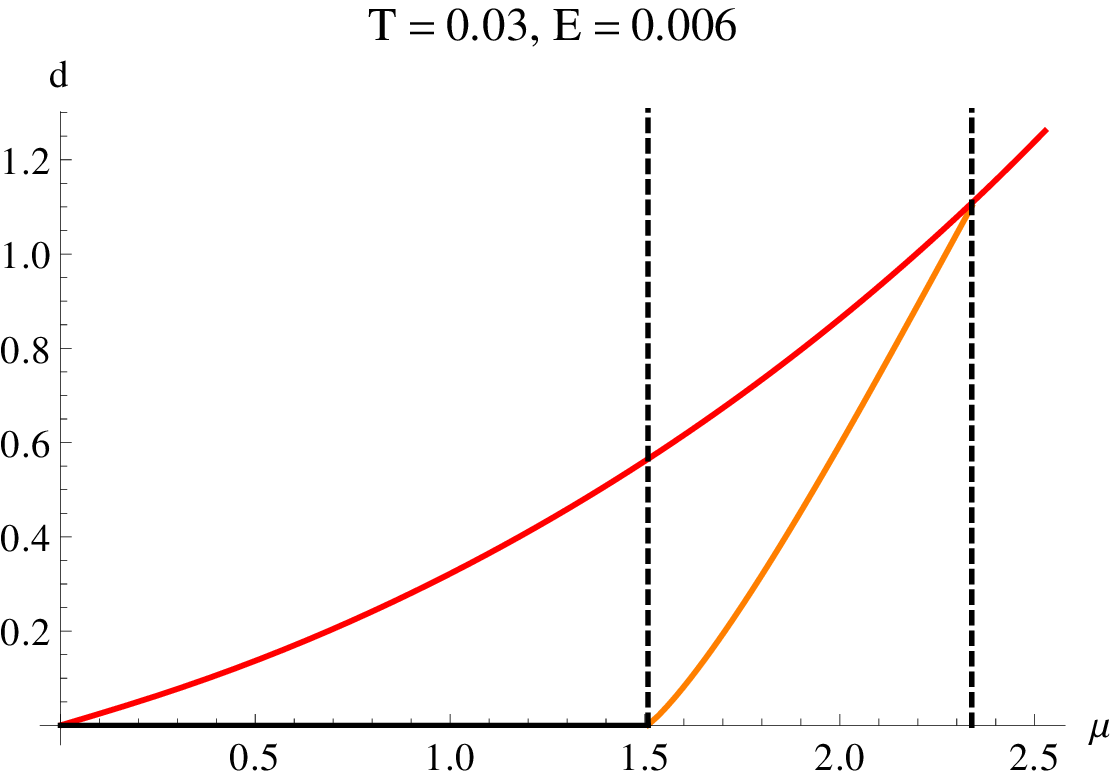} }
        \subfigure[2nd  $\ra$ 1st  $\ra$ 2nd order]  
   {\includegraphics[width=4.5cm]{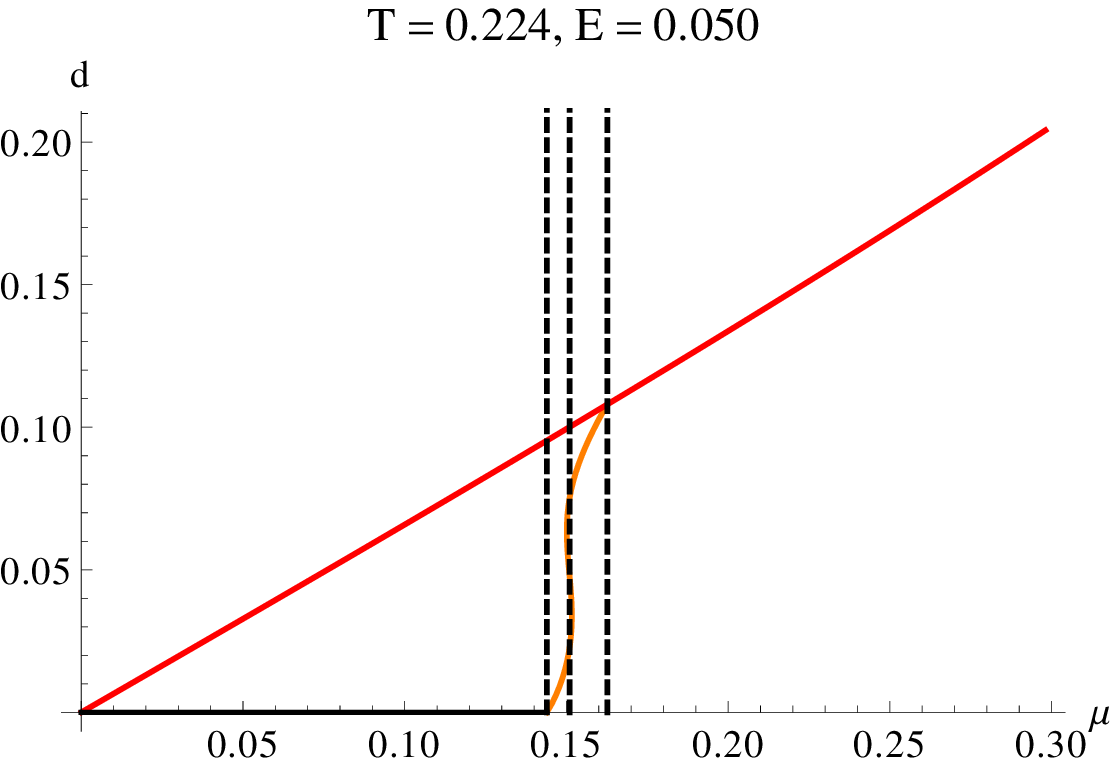}\label{dmud}}
           \subfigure[1st order]  
   {\includegraphics[width=4.5cm]{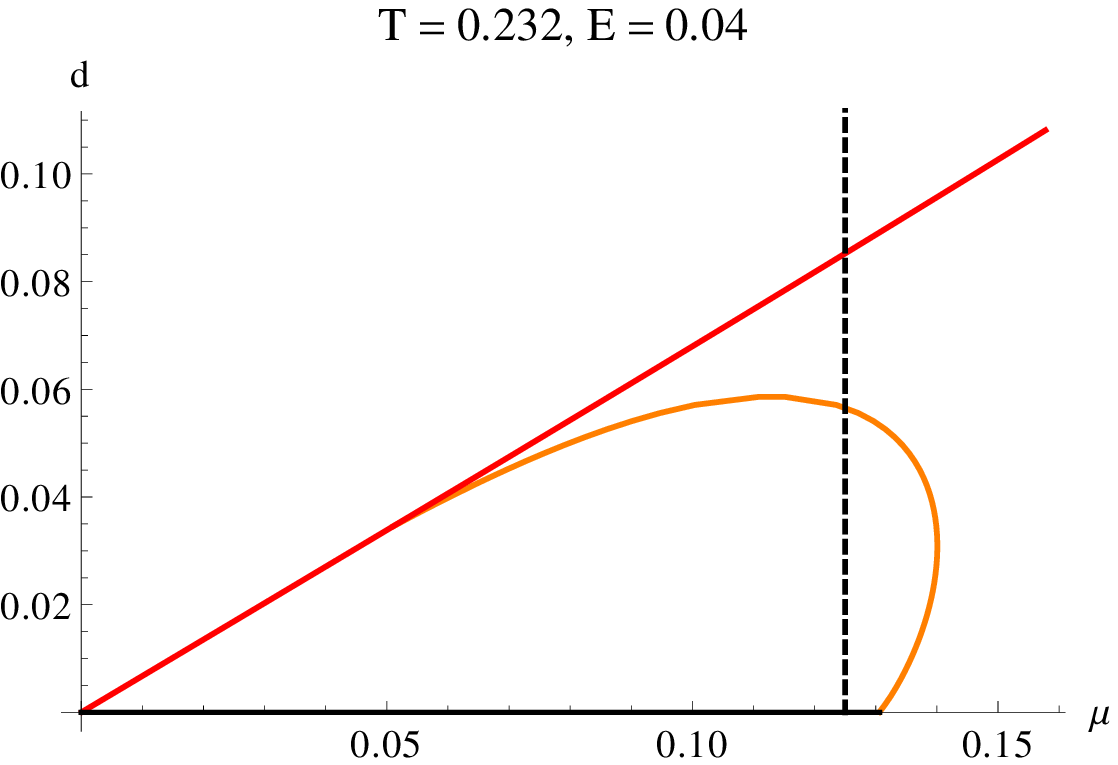}}
           \subfigure[2nd  order]  
   {\includegraphics[width=4.5cm]{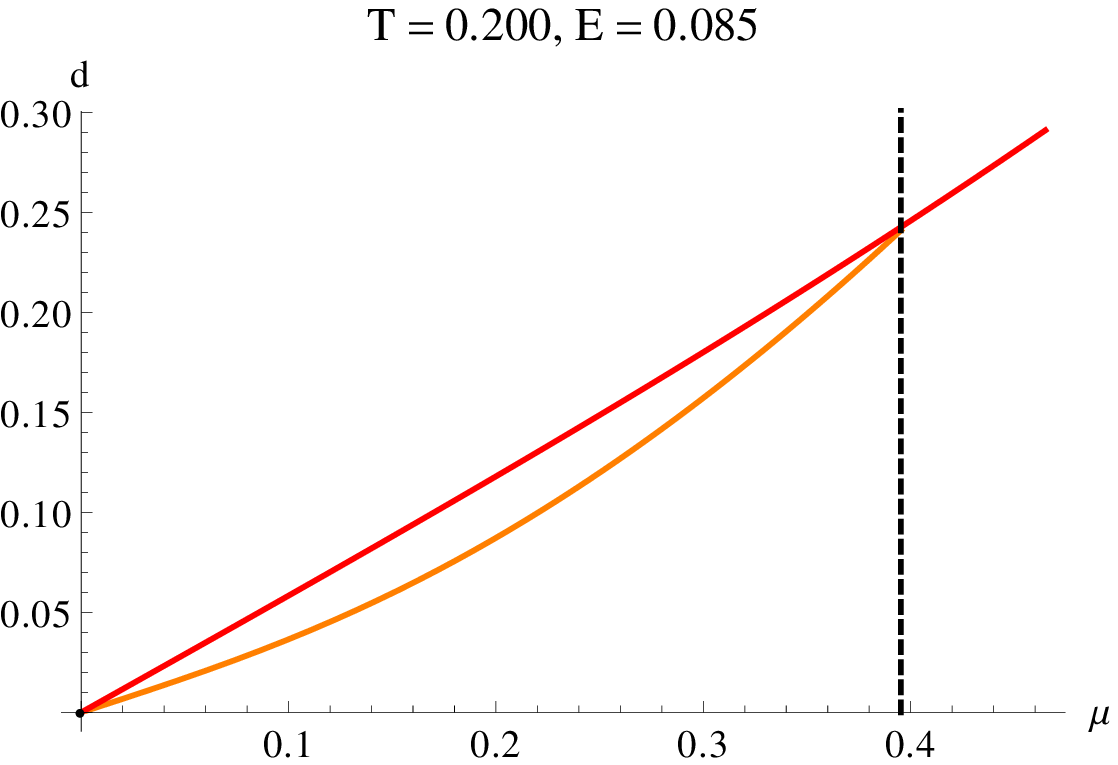}}
  \caption{
            Representative $\mu$-$d$ plots. The black line on the $\mu$ axis corresponds to
             a embedding that misses the singular shell; 
             the orange curve to a spike embedding that ends on the singular shell; 
             and the red curve to a flat embedding.  
             The transition points are shown by the vertical dotted lines. Subcaptions 
             are the order of transition as $\mu$ increases.
            }\label{dmu}
\end{figure}
\begin{figure}[]
\centering
\subfigure[$T=0.232$]
  {\includegraphics[width=4.5cm]{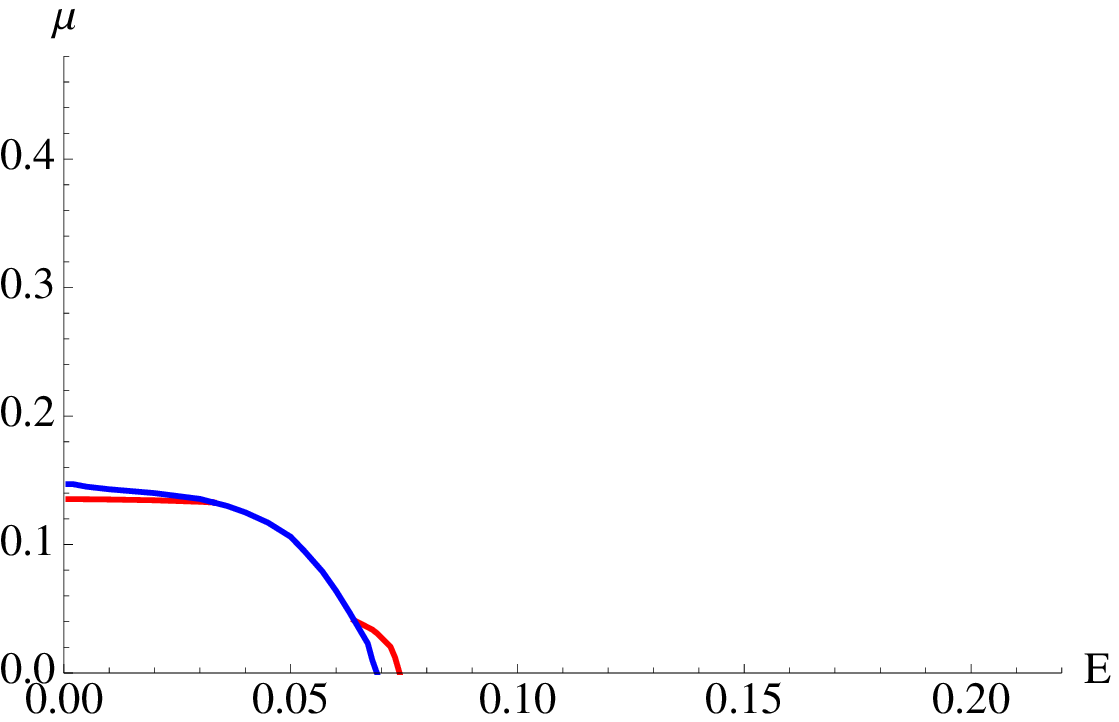}}
  \subfigure[$T=0.224$]
   {\includegraphics[width=4.5cm]{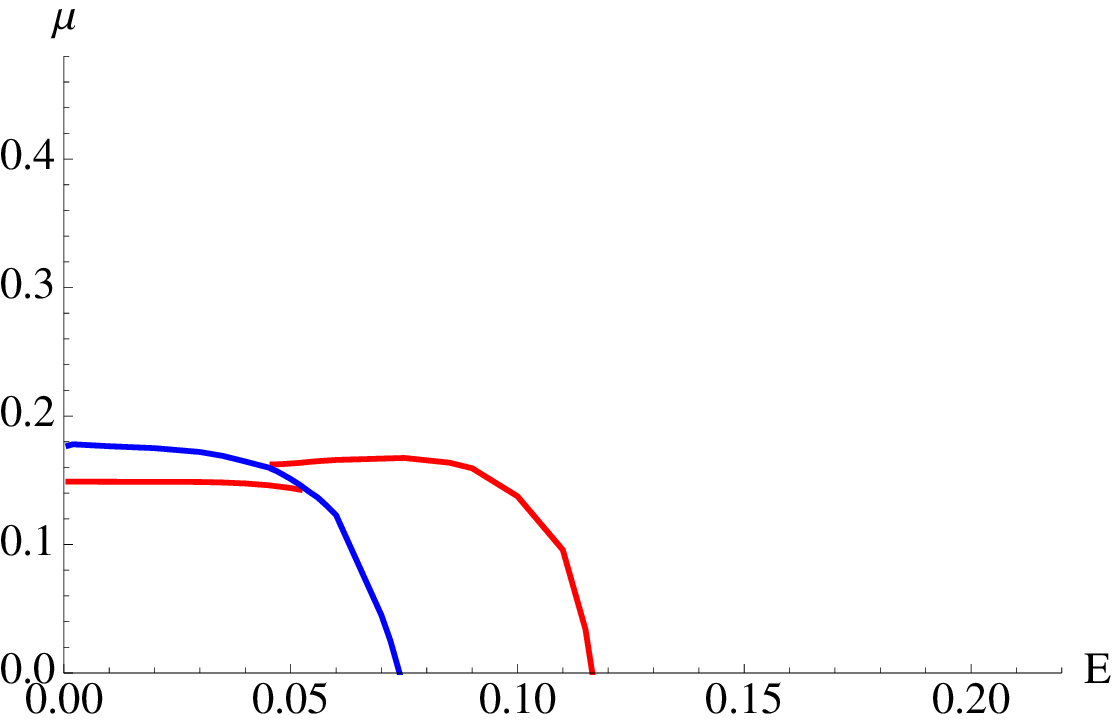}}
   \subfigure[$T=0.222$]
   {\includegraphics[width=4.5cm]{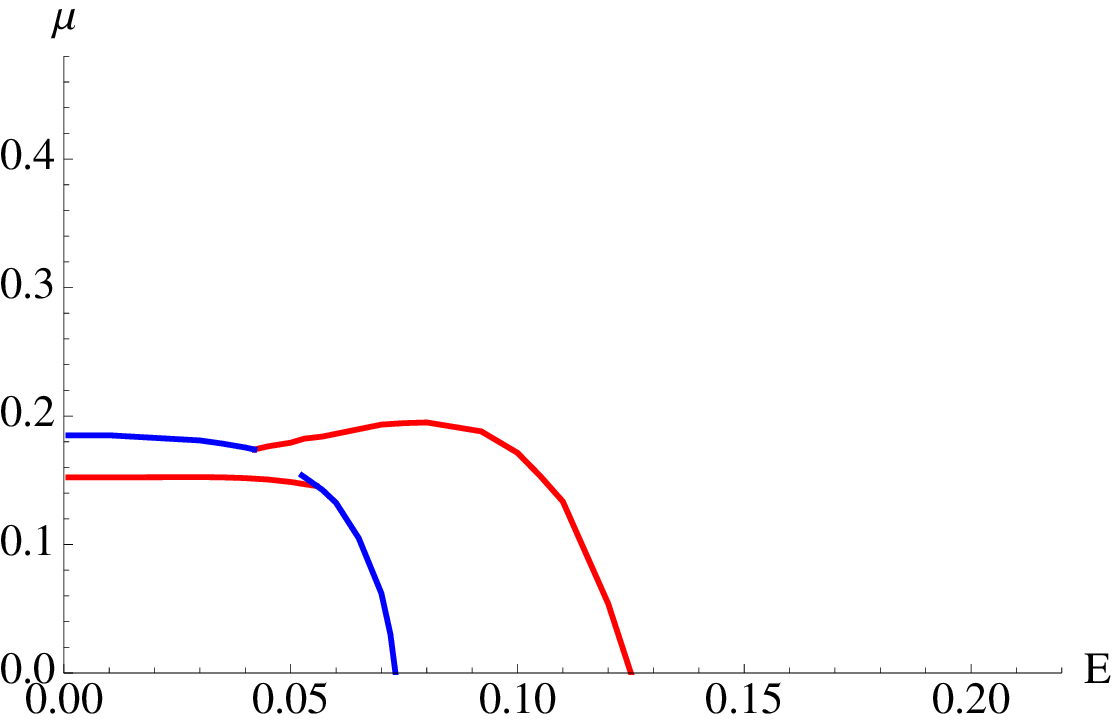}}
   \subfigure[$T=0.210$]
   {\includegraphics[width=4.5cm]{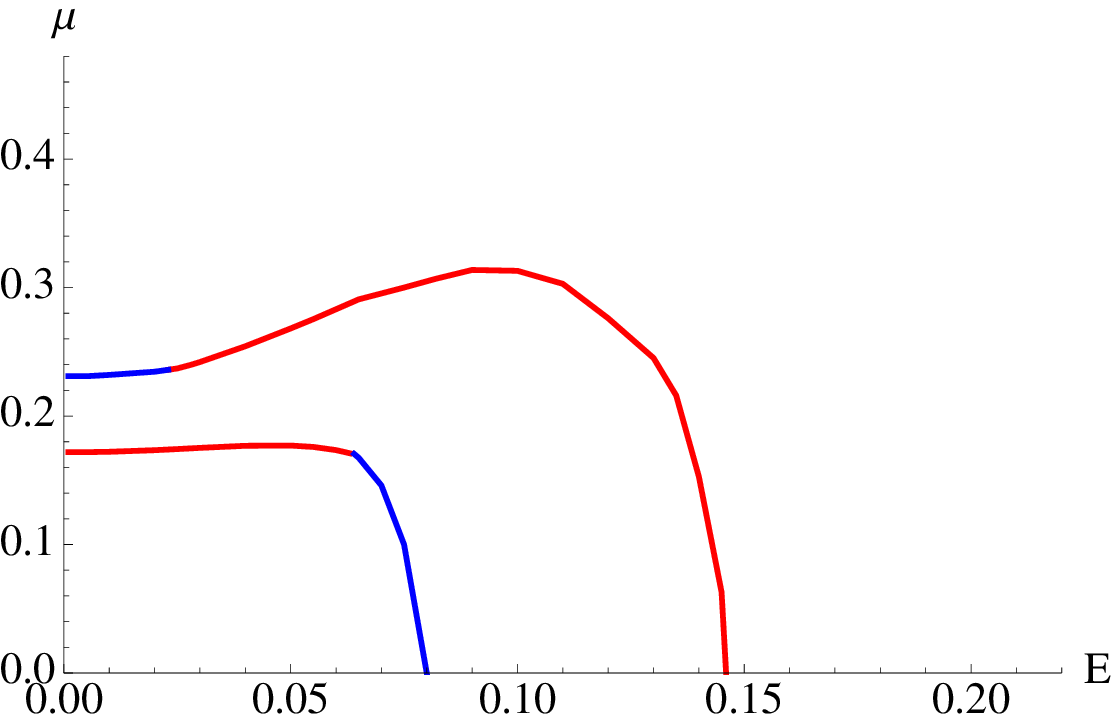}}
   \subfigure[$T=0.200$]
   {\includegraphics[width=4.5cm]{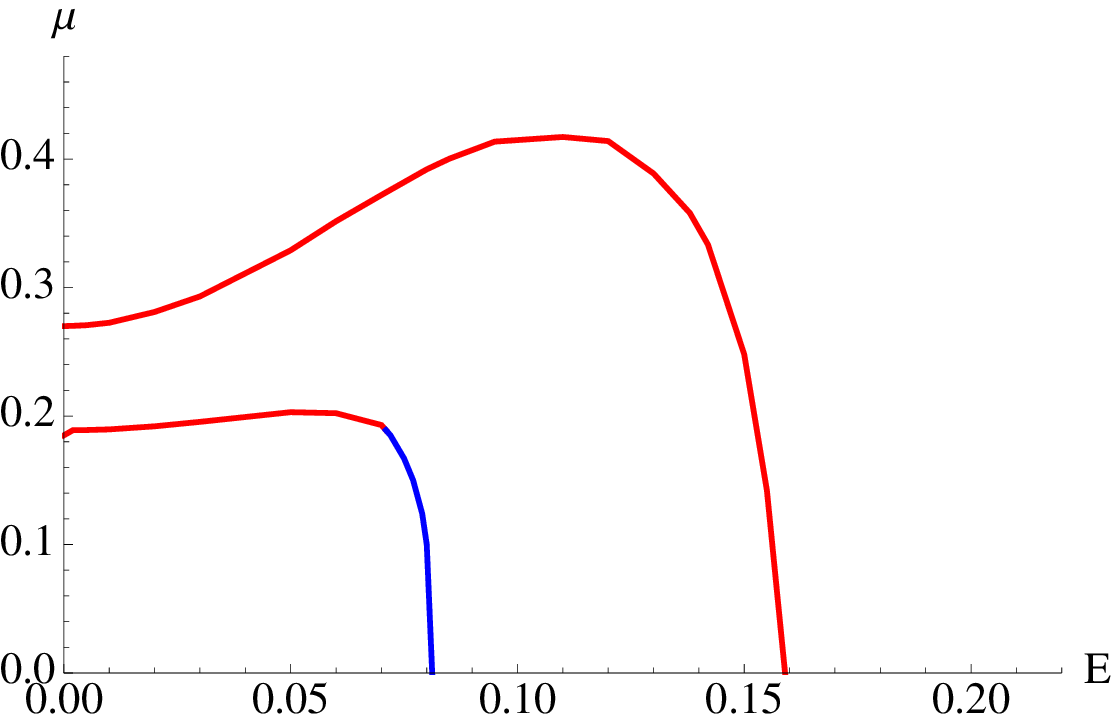}}
   \subfigure[$T=0.003$ (Notice that the scale of $\mu$ is $100$ times larger)]
   {\includegraphics[width=4.5cm]{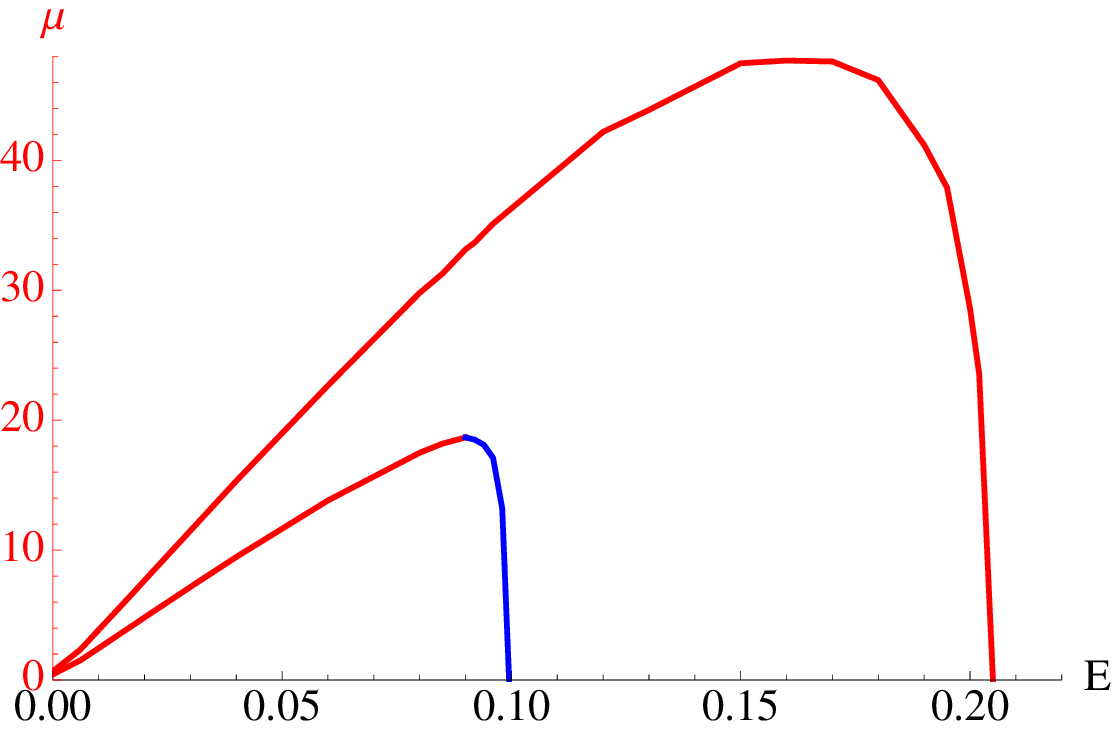}}
  \caption{ 
           { 
             Phase diagrams in the $\mu-E$ plane at various values of temprature 
             showing the phase structure. The solid blue lines are first order transitions, the 
             red lines are second order. }
           }\label{FixedT}
\end{figure}

In Fig \ref{dmu} we show sample plots from which each of a first and 
second order transition can be identified. 
There are in total six types of $d$-$\mu$ plots.
In each of the figures the black line on the $\mu$ axis corresponds to a chiral symmetry breaking
embedding that misses the singular shell; the orange curve to a spike embedding
that ends on the singular shell; and the red curve to a flat embedding.
It can be seen from the plots whether each transition between embeddings is smooth
and hence second order, or whether there is an `S' shaped structure so that
one expects a first order transition. The transition points are shown
by the vertical dotted lines. 
We have used these techniques on constant $E$ lines on each constant $T$ plane 
to determine the transition places and orders. Fig \ref{FixedT} 
is constructed from sequences of $d$-$\mu$ plots (Fig \ref{dmu}). 
For example, Fig \ref{FixedT} shows the following evolution of the $d$-$\mu$ plot as $E$ increases. 
\begin{eqnarray}
  \mathrm{Fig}\ \ref{FixedT}& :& \qquad \qquad \mathrm{Fig}\ \ref{dmu} \nonumber \\
  (a)&:& \quad (a)\ra(e)\ra(b)\ra(f) \nonumber \\
  (b)&:& \quad (a)\ra(d)\ra(b)\ra(f) \nonumber \\
  (c)&:& \quad (a)\ra(c)\ra(d)\ra(b)\ra(f) \nonumber \\
  (d)&:& \quad (a)\ra(c)\ra(b)\ra(f) \nonumber \\
  (e),(f)&:& \quad (c)\ra(b)\ra(f)
\end{eqnarray}

In Fig \ref{FixedT} we show six slices through the volume at varying $T$. 
Starting at high temperature the theory lives in the chirally symmetric phase 
with unstable mesons
and the material is a conductor. 
As the temperature falls to $T\simeq0.25$  the first order transition
to the chiral symmetry breaking, stable meson, insulator regime begins to appear in the $\mu$-$E$ plane
around $\mu=E=0$. That transition then expands away from the origin and remains briefly first order.

Our plot for $T=0.232$ shows the first interesting structure. Two areas in the plane grow out from 
the first order line bordered by additional second order transitions. In these areas the theory is 
in a chiral symmetry breaking but conducting
phase. The critical points where the first and second order transitions meet migrate inwards along the first
order boundary from each axis as the temperature falls and eventually pass each other as shown for $T=0.224$. 
At temperatures below that point 
there are constant $E$ trajectories across which there are three transitions. A second
order transition from conductor to insulator, a first order transition between two spike embeddings
and finally a second order chiral symmetry restoration transition. An example of
the relevant density chemical potential plot for this case is in Fig \ref{dmud}.

Our plot for $T=0.222$ shows the next key transition. The first order line between the two spike
embeddings breaks making the spike embedding phase continuously connected although
there is a remnant of the first order transition ending free at a critical point.

The first order transition near the $E=0$ phase boundary then diminish
as temperature is further reduced retreating towards the axes - see the plot for $T=0.21$.
It has totally disappeared by $T=0.2$. 
However the first order transition element on the conductor insulator
transition grows from the $E$ axis as temperature decrease. 

The final interesting feature begins to appear in the plot at $T=0.21$ where the phase boundaries
have begun to deform. They expands out to large $\mu$ very rapidly. Note that 
the scale of the $\mu$ axis at $T=0.03$ is $100$ times larger than the others. If we drew it 
on the same scale as the others it would look like Fig \ref{muvsE}: 
the $\mu$-independence of the $T=0$ limit is starting to be seen.
At zero $T$ (Fig \ref{muvsE}) there is a first order transition
between the $\mu$ axis and the rest of the $\mu, E$ plane - here we see that that forms as the
second order boundaries are pressed onto the $\mu$ axis. 

These results have been incorporated into the 3D plot in Fig~\ref{3D} which summarizes the full 
and rich phase structure.

\section{Quark Mass}

The analysis above has been purely for zero quark
mass. We have not performed an analysis of the introduction of
a quark mass, however, we note here that
an immediate consequence of introducing a small quark
mass is that the second order chiral symmetry restoration
transitions (the outer red lines in our figures) become
crossovers. We would expect some remnant of the
first order segment of this transition to remain associated
with a transition at which there is a discontiunuity in the
quark condensate even in the infinite mass limit where
the theory becomes the massive $\caln = 2$ theory
(we observed this in the $E=0$ limit in~\cite{Evans1}).
The insulator conductor transition remains with quark
mass and will again become that of the $\caln = 2$ theory at
large mass. At finite mass and temperature but without magnetic field, 
the insulator-conductor transition was shown to be first order in~\cite{Veselin1,Erdmenger1}.

\section{Summary}

We have explored the phase structure of the ${\cal N}=2$ gauge theory whose dual is the 
D3/D7 system. A magnetic field, $B$, tries to induce chiral symmetry breaking. An electric field, $E$, 
tries to disassociate the mesons of the theory and makes it a conductor. 
Finite density, $d$, (or chemical potential $\mu$)
and temperature, T, each favour melting of the mesons. The competition between these effects 
lead to a rich phase structure.

In Fig \ref{Tvsmu} we display the $T,\mu$ phase plane for the 
massless theory at fixed magnetic field. There are three phases -
at low $T,\mu$ a chiral symmetry breaking phase with stable mesons;
at intermediate values a chiral symmetry breaking phase with unstable 
mesons; and at large $T,\mu$ a chirally symmetric phase with unstable mesons. The transitions between these 
are a mix of first and second order transitions linked at two critical points. 

In Fig \ref{TvsE} we show the $T,E$ phase plane for the 
massless theory at fixed magnetic field. There are again three phases - 
at low $T,E$ a chiral symmetry breaking phase with stable mesons which acts as an insulator in
the presence of a small  electric field; at intermediate values a chiral symmetry breaking phase with unstable 
mesons which is a conductor and
sustains a current in the presence of an electric field; and at high $T,E$ a conducting 
chirally symmetric phase with unstable mesons. The transitions between these 
are again a mix of first and second order transitions linked at one critical point. 

The $E,\mu$ phase plane has rather different structure (Fig \ref{muvsE}). In particular for any finite $E$ 
the plane is $\mu$ independent and density is zero. At low $E$ we have a chiral symmetry breaking, insulator phase; at
intermediate $E$ a chiral symmetry breaking but conducting phase; at large $E$ a conducting and chirally 
symmetric phase. That the presence of infintessimal electric field 
does not allow density even with a very large chemical potential
and stops the restoration of chiral symmetry is rather surprising. The conclusion is
certainly correct within the analysis that we have performed.  It is
possible though that additional stringy physics should be present near the IR boundary 
in this limit to explain the sink for the energy the $E$ field is injecting. Such physics could
potentially change the phase structure at low $T$. 

Finally we have explored the full $E,\mu,T$ volume at fixed $B$ to show how these phases are
linked. The phase diagram is summarized  in Fig \ref{3D} with the transition boundaries 
marked. 

The variety of phase structure and transition type in such a simple theory is remarkable.
Whether these results can serve as an exemplar for other gauge theories either qualitatively or 
quantitatively remains to be seen but they certainly suggest a rich structure of phases
will be present in many gauge theories. 

{\bf Acknowledgements:} NE is grateful for the support of
an STFC rolling grant.  KK and AG are grateful for University of Southampton
Scholarships. We would like to thank 
Jonathan Shock, Javier Tarrio, Andy O'Bannon, and Veselin Filev 
for discussions.


\begin{thebibliography}{ll}


\bibitem{Malda}
J.~M.~Maldacena,  Adv.\ Theor.\ Math.\ Phys.\  {\bf 2}, 231 (1998)
Int.\ J.\ Theor.\ Phys.\  {\bf 38}, 1113 (1999)
[arXiv:hep-th/9711200].
%%CITATION = HEP-TH 9711200;%%

\bibitem{Witten:1998qj}
  E.~Witten,
%  ``Anti-de Sitter space and holography,''
  Adv.\ Theor.\ Math.\ Phys.\  {\bf 2} (1998) 253
  [arXiv:hep-th/9802150].
  %%CITATION = 00203,2,253;%%


\bibitem{Gubser:1998bc}
  S.~S.~Gubser, I.~R.~Klebanov and A.~M.~Polyakov,
%  ``Gauge theory correlators from non-critical string theory,''
  Phys.\ Lett.\  B {\bf 428} (1998) 105
  [arXiv:hep-th/9802109].
  %%CITATION = PHLTA,B428,105;%%


\bibitem{Karch}
  A.~Karch and E.~Katz,
%  ``Adding flavor to AdS/CFT,''
  JHEP {\bf 0206}, 043 (2002)
  [arXiv:hep-th/0205236].
  %%CITATION = JHEPA,0206,043;%%

\bibitem{Polchinski}
  M. Grana and J. Polchinski,
%  ``Gauge-gravity duals with holomorphic dilaton,''
  Phys. Rev. {\bf D65} (2002) 126005,
  [arXiv: hep-th/0106014].

\bibitem{Bertolini:2001qa}
  M.~Bertolini, P.~Di Vecchia, M.~Frau, A.~Lerda and R.~Marotta,
%  ``N = 2 gauge theories on systems of fractional D3/D7 branes,''
  Nucl.\ Phys.\  B {\bf 621}, 157 (2002)
  [arXiv:hep-th/0107057].
  %%CITATION = NUPHA,B621,157;%%

\bibitem{Mateos}
 M.~Kruczenski, D.~Mateos, R.~C.~Myers and D.~J.~Winters,
% ``Meson spectroscopy in AdS-CFT with flavor,''
 JHEP {\bf 0307} 049, 2003
 [arXiv:hep-th/0304032].
 %%CITATION = HEP-TH 0304032;%%

%\cite{Erdmenger:2007cm}
\bibitem{Erdmenger:2007cm}
  J.~Erdmenger, N.~Evans, I.~Kirsch and E.~Threlfall,
%  ``Mesons in Gauge/Gravity Duals - A Review,''
  Eur.\ Phys.\ J.\  A {\bf 35} (2008) 81
  [arXiv:0711.4467 [hep-th]].
  %%CITATION = EPHJA,A35,81;%%

\bibitem{Johnson1}
  V.~G.~Filev, C.~V.~Johnson, R.~C.~Rashkov and K.~S.~Viswanathan,
% ``Flavoured large N gauge theory in an external magnetic field,''
  JHEP {\bf 0710}, 019 (2007)
  [arXiv:hep-th/0701001];
  %%CITATION = JHEPA,0710,019;%%
%\bibitem{Albash:2007bk}
  T.~Albash, V.~G.~Filev, C.~V.~Johnson and A.~Kundu,
  %``Finite Temperature Large N Gauge Theory with Quarks in an External Magnetic
  %Field,''
  JHEP {\bf 0807}, 080 (2008)
  [arXiv:0709.1547 [hep-th]];
%\cite{Filev:2009xp}
%\bibitem{Filev:2009xp}
  V.~G.~Filev, C.~V.~Johnson and J.~P.~Shock,
  %``Universal Holographic Chiral Dynamics in an External Magnetic Field,''
  JHEP {\bf 0908} (2009) 013
  [arXiv:0903.5345 [hep-th]].
  %%CITATION = JHEPA,0908,013;%%
  


\bibitem{Evans2}
  N.~Evans, A.~Gebauer, K.~Y.~Kim and M.~Magou,
  %``Phase diagram of the D3/D5 system in a magnetic field and a BKT
  %transition,''
  arXiv:1003.2694 [hep-th].
  %%CITATION = ARXIV:1003.2694;%%

%\cite{Jensen:2010ga}
\bibitem{Jensen}
  K.~Jensen, A.~Karch, D.~T.~Son and E.~G.~Thompson,
  %``Holographic Berezinskii-Kosterlitz-Thouless Transitions,''
  Phys.\ Rev.\ Lett.\  {\bf 105}, 041601 (2010)
  [arXiv:1002.3159 [hep-th]].
  %%CITATION = PRLTA,105,041601;%%


%\cite{Evans:2010np}
\bibitem{Evans3}
  N.~Evans, K.~Jensen and K.~Y.~Kim,
  %``Non Mean-Field Quantum Critical Points from Holography,''
  Phys.\ Rev.\  D {\bf 82}, 105012 (2010)
  [arXiv:1008.1889 [hep-th]].
  %%CITATION = PHRVA,D82,105012;%%





%\cite{Evans:2010iy}
\bibitem{Evans1}
  N.~Evans, A.~Gebauer, K.~Y.~Kim and M.~Magou,
  %``Holographic Description of the Phase Diagram of a Chiral Symmetry Breaking
  %Gauge Theory,''
  JHEP {\bf 1003}, 132 (2010)
  [arXiv:1002.1885 [hep-th]].
  %%CITATION = JHEPA,1003,132;%%
%\cite{Evans:2010hi}
  
  
  

  
  \bibitem{Babington}
  J.~Babington, J.~Erdmenger, N.~J.~Evans, Z.~Guralnik and I.~Kirsch,
%  ``Chiral symmetry breaking and pions in non-supersymmetric gauge /  gravity
%  duals,''
  Phys.\ Rev.\  D {\bf 69} (2004) 066007
  [arXiv:hep-th/0306018]; T.~Albash, V.~G.~Filev, C.~V.~Johnson and A.~Kundu,
%  ``A topology-changing phase transition and the dynamics of flavour,''
  Phys.\ Rev.\  D {\bf 77} (2008) 066004
  [arXiv:hep-th/0605088]; D.~Mateos, R.~C.~Myers and R.~M.~Thomson,
%  ``Holographic phase transitions with fundamental matter,''
  Phys.\ Rev.\ Lett.\  {\bf 97} (2006) 091601
  [arXiv:hep-th/0605046];
  D.~Mateos, R.~C.~Myers and R.~M.~Thomson,
%  ``Thermodynamics of the brane,''
  JHEP {\bf 0705}, 067 (2007)
  [arXiv:hep-th/0701132].
  %%CITATION = JHEPA,0705,067;%%
  
  

%\cite{Kobayashi:2006sb}
\bibitem{Myers1}
%\bibitem{Sin1}
  S.~Nakamura, Y.~Seo, S.~J.~Sin and K.~P.~Yogendran,
%  ``A new phase at finite quark density from AdS/CFT,''
  J.\ Korean Phys.\ Soc.\  {\bf 52}, 1734 (2008)
  [arXiv:hep-th/0611021];
  %%CITATION = JKPSD,52,1734;%%
  S.~Kobayashi, D.~Mateos, S.~Matsuura, R.~C.~Myers and R.~M.~Thomson,
%  ``Holographic phase transitions at finite baryon density,''
  JHEP {\bf 0702}, 016 (2007)
  [arXiv:hep-th/0611099];
  %%CITATION = JHEPA,0702,016;%%  
%\cite{Nakamura:2007nx}
%\bibitem{Sin2}
  S.~Nakamura, Y.~Seo, S.~J.~Sin and K.~P.~Yogendran,
%  ``Baryon-charge Chemical Potential in AdS/CFT,''
  Prog.\ Theor.\ Phys.\  {\bf 120}, 51 (2008)
  [arXiv:0708.2818 [hep-th]];
  %%CITATION = PTPKA,120,51;%%
%\cite{Mateos:2007vc}
%\bibitem{Myers3}
  D.~Mateos, S.~Matsuura, R.~C.~Myers and R.~M.~Thomson,
%  ``Holographic phase transitions at finite chemical potential,''
  JHEP {\bf 0711} (2007) 085
  [arXiv:0709.1225 [hep-th]].
  %%CITATION = JHEPA,0711,085;%%
 
  
  %\cite{Grosse:2007ty}
\bibitem{Grosse:2007ty}
  J.~Grosse, R.~A.~Janik and P.~Surowka,
  %``Flavors in an expanding plasma,''
  Phys.\ Rev.\  D {\bf 77} (2008) 066010
  [arXiv:0709.3910 [hep-th]].
  %%CITATION = PHRVA,D77,066010;%%



%\cite{Evans:2010xs}
\bibitem{Evans4}
  N.~Evans, T.~Kalaydzhyan, K.~Y.~Kim and I.~Kirsch,
  %``Non-equilibrium physics at a holographic chiral phase transition,''
  JHEP {\bf 1101}, 050 (2011)
  [arXiv:1011.2519 [hep-th]].
  %%CITATION = JHEPA,1101,050;%%



%\cite{Guralnik:2011xz}
\bibitem{Guralnik:2011xz}
  G.~Guralnik, Z.~Guralnik and C.~Pehlevan,
  %``Dynamics of the chiral phase transition from AdS/CFT duality,''
  arXiv:1101.3095 [hep-th].
  %%CITATION = ARXIV:1101.3095;%%



%\cite{Evans:2010tf}
\bibitem{Evans5}
  N.~Evans, J.~French and K.~Y.~Kim,
  %``Holography of a Composite Inflaton,''
  JHEP {\bf 1011}, 145 (2010)
  [arXiv:1009.5678 [hep-th]].
  %%CITATION = JHEPA,1011,145;%%


%\cite{Karch:2007pd}
\bibitem{Andy1}
  A.~Karch and A.~O'Bannon,
  %``Metallic AdS/CFT,''
  JHEP {\bf 0709}, 024 (2007)
  [arXiv:0705.3870 [hep-th]].
  %%CITATION = JHEPA,0709,024;%%

%\cite{O'Bannon:2007in}
\bibitem{Hall}
  A.~O'Bannon,
  %``Hall Conductivity of Flavor Fields from AdS/CFT,''
  Phys.\ Rev.\  D {\bf 76}, 086007 (2007)
  [arXiv:0708.1994 [hep-th]].
  %%CITATION = PHRVA,D76,086007;%%


%\cite{Erdmenger:2007bn}
\bibitem{Erdmenger1}
  J.~Erdmenger, R.~Meyer and J.~P.~Shock,
%  ``AdS/CFT with Flavour in Electric and Magnetic Kalb-Ramond Fields,''
  JHEP {\bf 0712}, 091 (2007)
  [arXiv:0709.1551 [hep-th]].
  %%CITATION = JHEPA,0712,091;%%
%\cite{Albash:2007bq}



\bibitem{Veselin1}
  T.~Albash, V.~G.~Filev, C.~V.~Johnson and A.~Kundu,
  %``Quarks in an External Electric Field in Finite Temperature Large N Gauge
  %Theory,''
  JHEP {\bf 0808}, 092 (2008)
  [arXiv:0709.1554 [hep-th]].
  %%CITATION = JHEPA,0808,092;%%


%\cite{Karch:2008uy}
\bibitem{Andy2}
  A.~Karch, A.~O'Bannon and E.~Thompson,
  %``The Stress-Energy Tensor of Flavor Fields from AdS/CFT,''
  JHEP {\bf 0904}, 021 (2009)
  [arXiv:0812.3629 [hep-th]].
  %%CITATION = JHEPA,0904,021;%%


%\cite{Mas:2009wf}
\bibitem{Mas}
  J.~Mas, J.~P.~Shock and J.~Tarrio,
  %``Holographic Spectral Functions in Metallic AdS/CFT,''
  JHEP {\bf 0909}, 032 (2009)
  [arXiv:0904.3905 [hep-th]].
  %%CITATION = JHEPA,0909,032;%%
  
 
 %\cite{Ammon:2009jt}
\bibitem{Andy3}
  M.~Ammon, T.~H.~Ngo and A.~O'Bannon,
  %``Holographic Flavor Transport in Arbitrary Constant Background Fields,''
  JHEP {\bf 0910}, 027 (2009)
  [arXiv:0908.2625 [hep-th]].
  %%CITATION = JHEPA,0910,027;%% 


\bibitem{KST}
  K.~Y.~Kim, J.~P.~Shock and J.~Tarrio,
  %``The open string membrane paradigm with external electromagnetic fields,''
  arXiv:1103.4581 [hep-th].
  %%CITATION = ARXIV:1103.4581;%%


%\cite{Peeters:2006iu}
\bibitem{Peeters:2006iu}
  K.~Peeters, J.~Sonnenschein and M.~Zamaklar,
%  ``Holographic melting and related properties of mesons in a quark gluon
%  plasma,''
  Phys.\ Rev.\  D {\bf 74} (2006) 106008
  [arXiv:hep-th/0606195];
  C.~Hoyos-Badajoz, K.~Landsteiner and S.~Montero,
%  ``Holographic Meson Melting,''
  JHEP {\bf 0704} (2007) 031
  [arXiv:hep-th/0612169].
  %%CITATION = JHEPA,0704,031;%%
  



%\cite{Apreda:2005yz}
\bibitem{Apreda:2005yz}
  R.~Apreda, J.~Erdmenger, N.~Evans and Z.~Guralnik,
  %``Strong coupling effective Higgs potential and a first order thermal  phase
  %transition from AdS/CFT duality,''
  Phys.\ Rev.\  D {\bf 71}, 126002 (2005)
  [arXiv:hep-th/0504151].
  %%CITATION = PHRVA,D71,126002;%%


%\cite{Kim:2007zm}
%\bibitem{Kim:2006gp}
\bibitem{Kim}
  K.~Y.~Kim, S.~J.~Sin and I.~Zahed,
%  ``Dense hadronic matter in holographic QCD,''
  arXiv:hep-th/0608046;
  %%CITATION = HEP-TH/0608046;%%
  K.~Y.~Kim, S.~J.~Sin and I.~Zahed,
%  ``The Chiral Model of Sakai-Sugimoto at Finite Baryon Density,''
  JHEP {\bf 0801}, 002 (2008)
  [arXiv:0708.1469 [hep-th]];
  %%CITATION = JHEPA,0801,002;%%
%\cite{Kim:2006gp}





%\cite{Kim:2008zn}
\bibitem{KE}
  K.~-Y.~Kim, S.~-J.~Sin, I.~Zahed,
  %``Dense and Hot Holographic QCD: Finite Baryonic E Field,''
  JHEP {\bf 0807}, 096 (2008).
  [arXiv:0803.0318 [hep-th]];
%\cite{Davody:2011yx}
%\bibitem{Davody:2011yx}
  A.~Davody,
  %``Noncritical Holographic QCD in External Electric Field,''
  [arXiv:1102.4509 [hep-th]].




\end{thebibliography}
\end{document}